\documentclass[aps,prb,twocolumn,superscriptaddress,floatfix,longbibliography]{revtex4-2}

\usepackage[colorlinks=true,citecolor=blue]{hyperref}
\usepackage{graphicx}
\usepackage{amsmath}
\usepackage{amssymb} 
\usepackage{hyperref}
\usepackage{float}
\usepackage[dvipsnames,svgnames,x11names,hyperref]{xcolor}
\DeclareGraphicsExtensions{.png,.jpg,.eps}
\usepackage{xcolor}

\newcommand{\B}{ \mathcal{J} }

\newcommand{\sign}{\text{sign}}
\begin{document}

\author{Pramod Kumar}
\affiliation{Department of Applied Physics, Aalto University, 00076 Aalto, Espoo, Finland}

\author{Guangze Chen}
\affiliation{Department of Applied Physics, Aalto University, 00076 Aalto, Espoo, Finland}

\author{J. L. Lado}
\affiliation{Department of Applied Physics, Aalto University, 00076 Aalto, Espoo, Finland}

\title{Kondo-lattice-mediated interactions in flat band systems}

\begin{abstract}
Electronic flat bands represent
a paradigmatic platform to realize strongly correlated matter due to their associated
divergent density of states. In common instances, including electron-electron interactions leads
to magnetic instabilities for repulsive interactions and superconductivity for attractive interactions.
Nevertheless, interactions of Kondo nature in flat band systems have remained relatively unexplored.
Here we address the emergence of interacting states mediated by Kondo lattice coupled to a flat band system.
Combining dynamical mean-field theory and tensor networks methods
to solve flat band Kondo lattice models in one and two dimensions, we show
the emergence of a robust underscreened regime leading
to a magnetically ordered state in the flat band.
Our results put forward flat band Kondo lattice models as a platform
to explore the genuine interplay between flat band physics and many-body Kondo screening.
\end{abstract}

\date{\today}

\maketitle
\section{Introduction}

Flat band systems represent one of the paradigmatic systems to engineer correlated matter~\cite{PhysRevLett.103.080406,PhysRevB.90.094506,PhysRevB.83.220503,PhysRevB.94.245149,PhysRevB.102.201112,PhysRevLett.126.027002}. 
Quantum engineering has provided a variety of platforms potentially combining
both flat bands and interactions, including atomic lattices\cite{Drost2017,Slot2017,PhysRevResearch.2.043426,Khajetoorians2019}, cold atoms\cite{PhysRevLett.108.045305,Taie2015,Bloch2012}
and twisted moire materials\cite{PhysRevB.82.121407,Cao2018super,Andrei2021}.
Their potential for correlated physics stems from the vanishing electronic dispersion,
which creates a greatly enhanced density of states at the Fermi energy~\cite{Leykam2018,Kopnin:2011,Heikkila:2011,doi:10.1142/S0217979215300078}. While a wide variety
of correlated states in flat band systems can emerge, minimal attractive or repulsive on-site
interactions are well known to lead to magnetism and superconductivity, respectively~\cite{PhysRevB.90.094506,Peotta2015,PhysRevA.90.043624,Cao2018,Cao20181,Yankowitz1059,PhysRevB.97.155125,PhysRevLett.127.026401}.
More complex interactions in flat bands are also well known to give rise to other symmetry broken
states, including charge density waves and bond orders~\cite{Jiang2019,PhysRevB.98.121406,Kumar2019,PhysRevA.103.L031301}. 
The different interactions considered
are usually written as an effective density-density interaction. Nevertheless, external couplings such as Kondo couplings~\cite{PhysRevB.29.3035,RevModPhys.73.797} in the system can lead to even more sophisticated interaction terms.

Electronic states coupled to magnetic impurities are known as prototypical many-body states\cite{Wirth2016,RevModPhys.69.809}.
The simplest example corresponds to the Kondo problem, in which a single magnetic impurity forms
a many-body ground state with its electronic bath~\cite{Wirth2016,RevModPhys.69.809,Si1161}. The lattice version of the problem, known as
the Kondo-lattice problem, represents the starting point for exotic physical phenomena found in heavy-fermion systems~\cite{Wirth2016,coleman2015heavy,PhysRevLett.112.116405,PhysRevB.93.035120,2021arXiv210311989V,PhysRevLett.104.106408,Dzero2016,PhysRevLett.115.156405,Jiao2020}.
Interestingly, the Kondo physics outlined above is usually addressed in systems with strong electronic dispersion,
while the Kondo problem for flat bands has been much less explored~\cite{PhysRevB.90.045112,PhysRevB.88.121107,PhysRevB.97.085123,PhysRevB.99.245118,PhysRevMaterials.3.084003}.
In the dispersive limit, the interaction between a local
magnetic impurity and the conduction bath is determined
by the Kondo temperature, increasing with the
density of states, and therefore divergent
in the flat band regime.
The coupling between magnetic impurities, known as Ruderman–Kittel–Kasuya–Yosida interaction\cite{PhysRev.96.99,PhysRev.106.893,Kasuya1956}, 
is determined by the Fermi wavelength
of the conduction bath. However, in the flat band limit, the previous picture breaks down due to the absence of a well defined Fermi surface.

\begin{figure}
\includegraphics[width=1.0\linewidth]{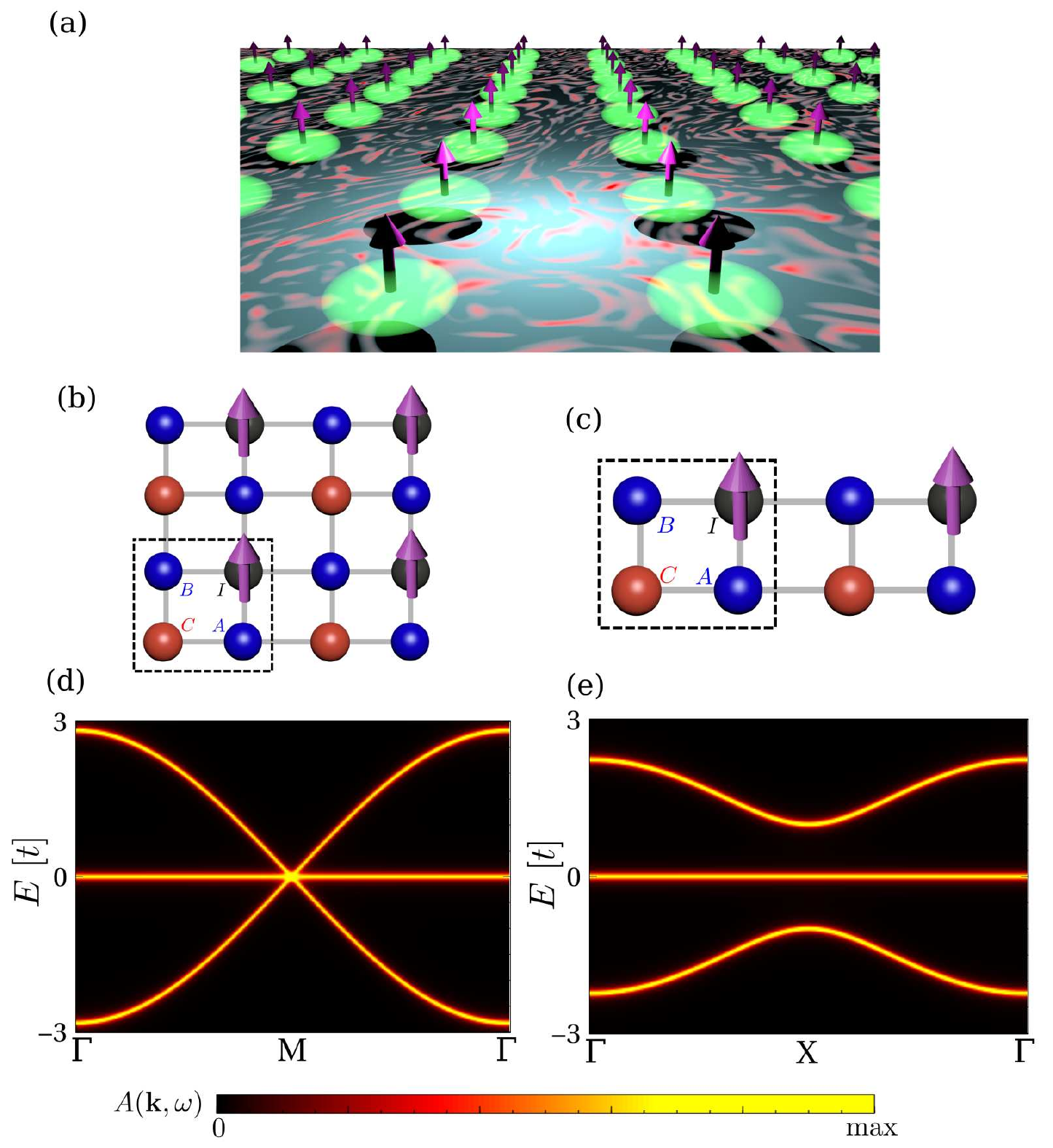}
\caption{(a) Sketch of a Kondo lattice coupled to a flat band electron gas.
Panels (b, c) show the lattice model realizing a flat band Kondo lattice model
in two dimensional (b) and one dimension (c).
Panels (d, e) show the momentum resolved spectral function $A(\mathbf k,\omega)$
of the flat band electron gas of panels (b, c) respectively.
}
\label{fig:lattice}
\label{fig:fig1}
\end{figure}

Here we address the fate of a flat band system coupled to a lattice of magnetic impurities, realizing the
so-called flat band Kondo lattice model. 
We observe that the system develops a robust
local magnetic order, overcoming Kondo
screening effects of the flat band.
We demonstrate that the full phenomenology can be
captured by symmetry broken
mean-field method,
and compare these results
with two genuine many-body
methods, dynamical mean-field theory and tensor networks states.
Our results demonstrate the non-trivial impact of flat bands in Kondo
lattice problems, emphasizing the complex interplay between exchange
and Kondo physics in flat band systems.

The manuscript is organized as follows.
In section~\ref{sec:model}, we introduce the minimal
models featuring a flat-band Kondo lattice physics, both in one dimension and two-dimensions.
In section \ref{sec:2d} we present the solution of the two-dimensional
flat band Kondo lattice model as solved by dynamical mean field theory.
In section \ref{sec:1d} we present the solution of the one-dimensional
flat band Kondo lattice model as solved with
tensor networks. Finally, in section \ref{sec:con} we summarize our conclusions.

\section{Flat band Kondo lattice model}
\label{sec:model}

In the following we describe the effective models
used to capture a flat band Kondo lattice model,
as shown in Fig.~\ref{fig:fig1} (bc). 
The full Hamiltonian 
$\mathcal H = 
\mathcal H_0 +
\mathcal H_{\text{int}}
$ has both non-interacting
$\mathcal H_0$
and interacting
$\mathcal H_{\text{int}}$
terms. Let us start with the
non-interacting term, for which we take a
first neighbor spinful model of the form
\begin{equation}
\mathcal{ H}_0 = t \sum_{\langle ij \rangle , \sigma} 
c^\dagger_{i \sigma} {c}^{\phantom{\dagger}}_{j \sigma}
\label{eq:h0}
\end{equation}

where 
$c_{i\sigma}$ is the annihilation operator
for site $i$ and spin $\sigma$, and
$\langle \rangle$ denotes nearest-neighbor sites. 
The lattices considered for Eq. \ref{eq:h0} would be a square lattice
and a ladder, for the two-dimensional and one-dimensional
cases, respectively.
In both cases, the fundamental unit cell of the system
consists of four spinful sites.

Let us now address the interacting
term of the Hamiltonian. In each unit cell of the system, we will
include interactions solely in a single site, which we label by $I$
in the form

\begin{equation}
\mathcal{  H}_{\rm int}=U\sum_{I} 
\left ( {n}_{I\uparrow}-\frac{1}{2} \right )
\left ( {n}_{I\downarrow}-\frac{1}{2}\right ),
\label{eq_Hu}
\end{equation}
where $ {n}_{I\sigma}=c_{I\sigma}^\dagger c^{\phantom{\dagger}}_{I\sigma}$ and $U>0$ is the interaction strength. 
The interaction acts only at one of every four sites,
namely the sites labelled as $I$, which in the following will be referred to as the impurity site. 
It is worth to note that by definition, the previous Hamiltonian leads to a half filled state
for arbitrary $U$.
In the strong coupling limit $U\gg t$,
the low energy sector realizes a flat band electron gas, as shown in Figs. \ref{fig:fig1}de. The interacting site
develops a local magnetic moment, leading to a low energy effective Hamiltonian for the full system
of the form

\begin{equation}
    \mathcal{  H}_{\text{eff}} 
    = 
    t \sum_{\langle ij \rangle', \sigma} 
c^\dagger_{i \sigma} {c}^{\phantom{\dagger}}_{j \sigma}
+
\sum_{\langle I j \rangle''}
J
\mathbf {S_I} \cdot \mathbf {\tau_{\sigma,\sigma'}} c^\dagger_{j,\sigma} c_{j,\sigma'} 
\label{eq:hk}
\end{equation}
where $\langle ij \rangle'$ denotes the
sum over first neighboring non-interacting sites,
and
$\langle I j \rangle''$ denotes the sum over first neighbors
with $I$ an impurity site and $j$ a non-interacting site.
The effective Kondo coupling $J$ depends on
the interaction $U$ as $J \sim t^2/U$. In the following we will
work with the full fermionic Hamiltonian $\mathcal H$, yet
the effective Hamiltonian as shown in Eq. \ref{eq:hk} will provide
useful insights to rationalize the full many-body solution.

Finally, it is worth to comment on the emergence of the flat band
in the previous models. In the infinite $U$ limit,
one of the four sites is fully disconnected from the
other non-interacting sites. In this limit,
the two-dimensional and one dimensional models
realize a bipartite lattice with a different number
of sublattice sites per unit cell,
automatically leading to a flat band.\cite{PhysRevB.34.5208,PhysRevB.66.014204,Mielke1991,PhysRevLett.62.1201}.
For finite $U$, the system will thus realize a two-dimensional or one-dimensional
electron gas coupled to a lattice of Kondo impurities.
In the following, we explore this finite-U coupling limit,
first in two dimensions with dynamical mean-field theory
and later in one dimension with tensor-networks.

\section{DMFT approach ot the two dimensional flat band Kondo lattice model}
\label{sec:2d}

In the following, we will address the fate of the two-dimensional Kondo lattice problem.
We will consider the mean-field solution as a reference, as to see how quantum
fluctuations included in the dynamical mean-field approach
renormalize the results.
 Evaluations of the site selective magnetism have been calculated using the cellular mean-field theory, cellular dynamical mean-field theory with continuous time quantum Monte-Carlo method, and exact diagonalization as the impurity solver~\cite{RevModPhys.68.13,PhysRevB.76.035116,Gull}. Within CDMFT, a lattice problem is mapped to a finite cluster coupled to a non-interacting bath. In our case, the cluster is a four site $(2\times2)$ plaquette as shown in Fig.~\ref{fig:lattice}~\cite{vanhala2017dynamical,PhysRevB.62.R9283,PhysRevB.74.054513}. We define the site-dependent
 absolute magnetization for the cluster as
$
m_{\text{i}}= \langle n_{i\uparrow}\rangle-
\langle n_{i\downarrow}\rangle 
$,
where $\langle n_{i,\sigma}\rangle=G_{i,\sigma\sigma}(\tau \rightarrow 0^-)$ is the density of spin-$\sigma$ particles for a given site of the cluster calculated from the local Green's function. To understand the origin of the site-selective magnetic order; we calculate the effective hybridization~\cite{PhysRevB.76.104425} between the impurity and the sites near it. We can define the nonlocal effective hybridization order for the given four site clusters as
\begin{eqnarray}
\Delta_\alpha=|\sum_{\alpha\neq \beta,\sigma}\langle c_{\alpha\sigma}^\dagger c_{\beta\sigma} \rangle|=\sum_{\alpha\neq \beta,\sigma}|G_{i,\sigma\sigma}^{\alpha\beta}(\tau \rightarrow 0^-)|\label{hyb}
\end{eqnarray} 
where $\alpha(\beta)$ is the sublattice index. The behavior of effective hybridization is proportional to the non-interacting local density of states at the sublattices of the unit cell~\cite{Si1161}.

\begin{figure}
\includegraphics[width=\columnwidth]{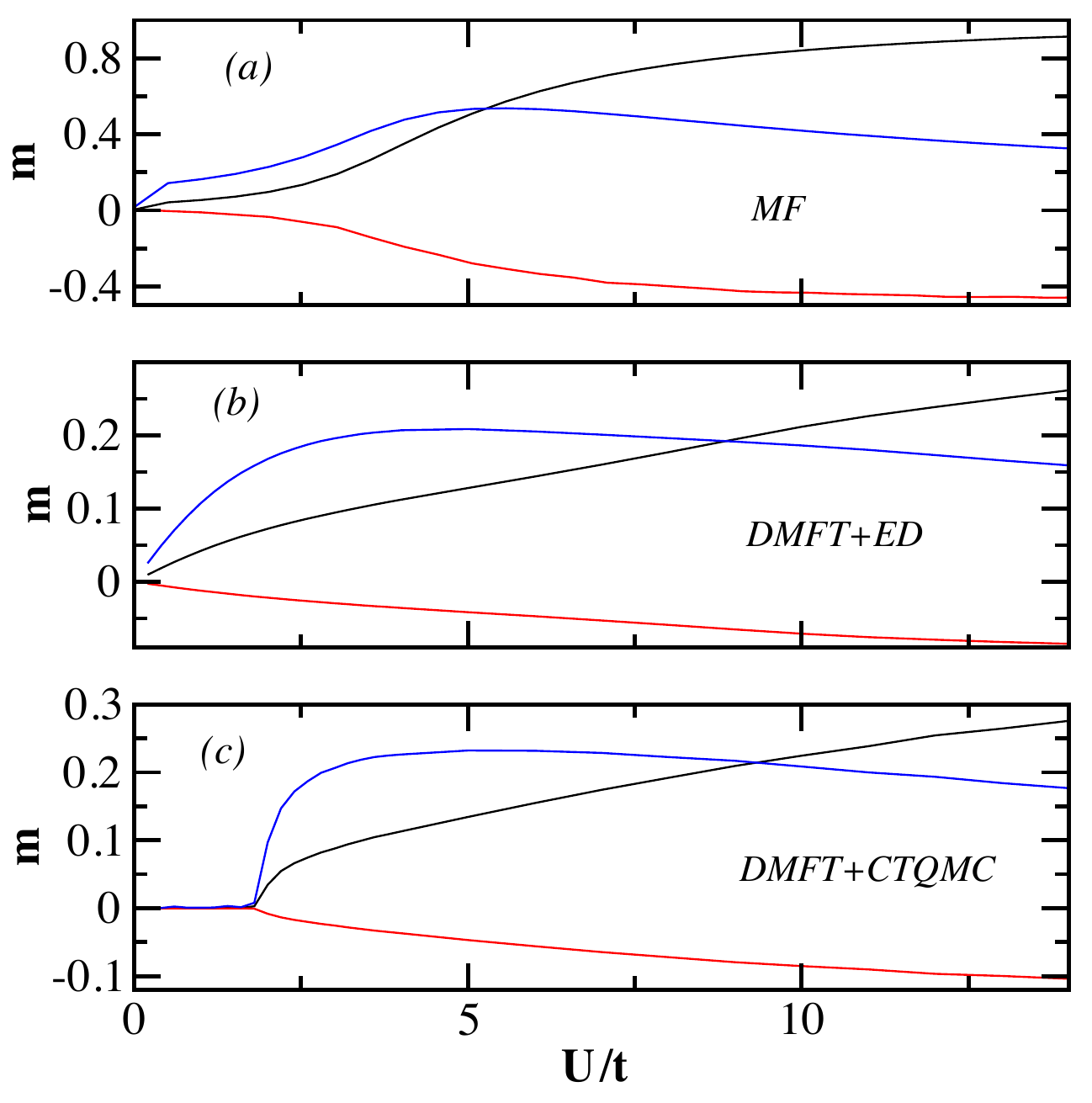}
\caption{Magnetic order parameter at impurity site, $I$ (black), at sites next to the impurity, $A/B$ (red), and at the second neighbor site of the impurity site, $C$ (blue), evaluated  
using (a) mean-field theory~\cite{PhysRevB.31.4403}, (b) dynamical mean-field theory (DMFT) + exact diagonalization (ED), and  (c) dynamical mean-field theory (DMFT) + continuous-time quantum Monte-Carlo (CTQMC) for varying interaction strength $U$. Note that the magnitude of the magnetic  order at $C$ site has been scaled by a factor of four for the purpose of visual clarity. }
\label{fig2:mag_order}
\end{figure}
In the following, we present the complete magnetic phase diagram in the presence of a finite two-body interaction $U$ at finite temperature $T$ at half-filling, where the number of particles per site is one. Simultaneously, we also explore the nature of effective Kondo-hybridization on the emergent site-selective magnetization. The role of quantum fluctuations, ignored in the mean field theory, have been addressed by using DMFT. 
 
\subsection{Zero temperature calculations}
\label{sec:magnetism at zero temperature}
\begin{figure}
\includegraphics[width=1.0\linewidth]{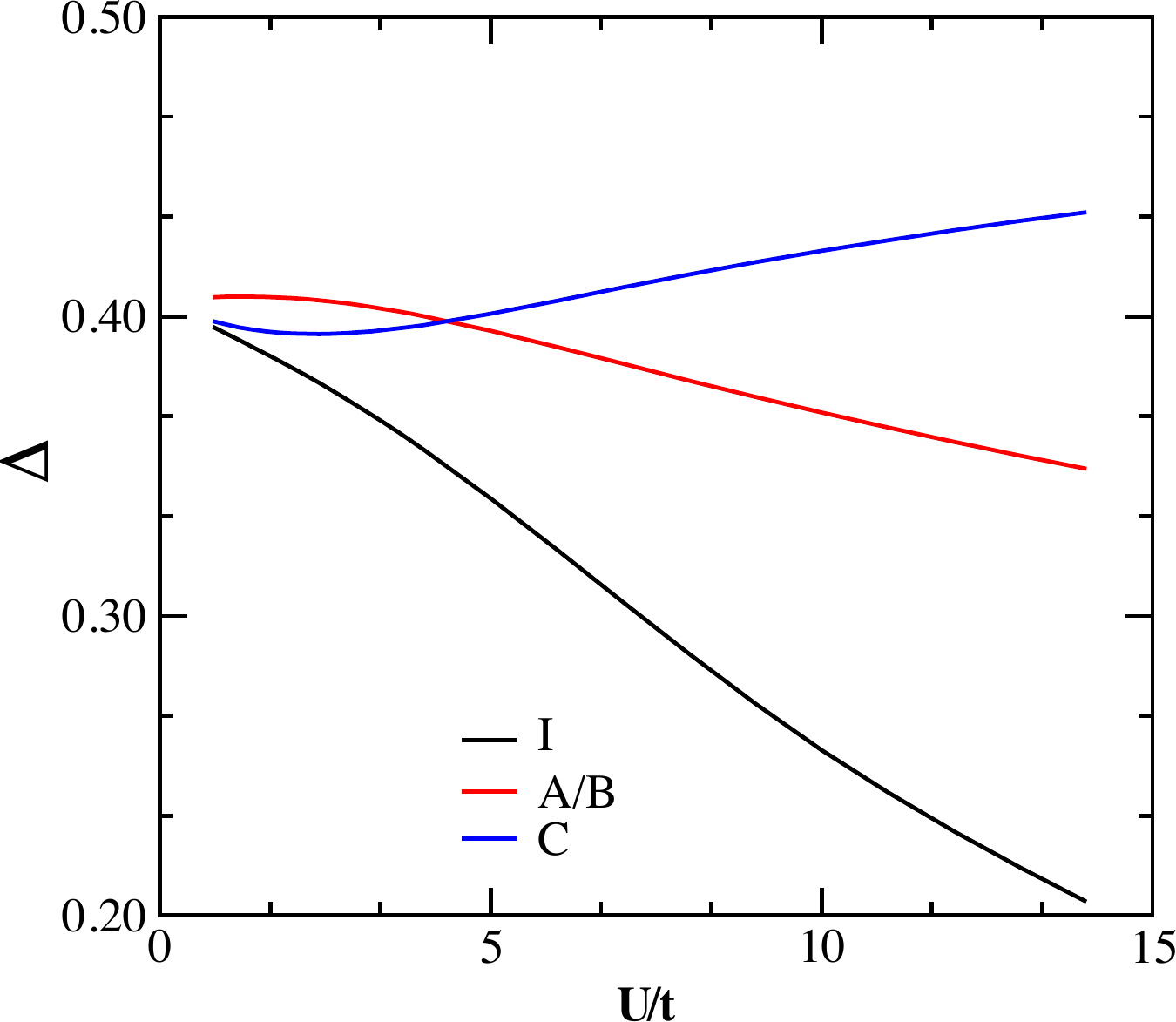}
\caption{Effective hybridization of a given sublattice with the rest of the sites in the unit cell as Eq~\ref{hyb}.}
\label{fig:hyb_gap}
\end{figure}

Due to emergent singularity in the LDOS at different sublattices in the presence of  the finite local interaction at the impurity site, the local magnetic order is non-uniform across different sites. We show the spatially resolved magnetic order $m_\alpha$ evaluated using Hartree-Fock mean field theory in the Fig.~\ref{fig2:mag_order}(a), zero temperature ED+CDMFT in the Fig.~\ref{fig2:mag_order}(b) and CTQMC+CDMFT at $T=0.01$  the Fig.~\ref{fig2:mag_order}(c) for varying interaction strength $U/t$ at the impurity site. We allow the breaking of the $SU(2)$ spin-rotation symmetry to capture the magnetically ordered state. An initial self-energy that is constant in the Matsubara frequency is added in this way that it breaks SU$(2)$ symmetry of the Hamiltonian.

For weak to moderate interactions, the local magnetic order at different sublattices gradually develops such that 
$\sign (m_{A/B})=-sign(m_{C})=-\sign(m_{I}$) for any $U/t > 0$ due to the antiferromagnetic
coupling between the local moments at neighboring site of the unit cell.
The sizes of the
magnetic moments of the $A/B$, $C$, and $I$ sites are very different,
as expected from the non-uniform nature of the model. 
While the magnetization of the $A/B$ and $I$ sites saturates with the strength of the interaction, the same does not occur on the site $C$. Magnetic order at the $C$ site asymptotically goes to zero after attaining the peak at a given $U/t$.
The qualitative behavior obtained from the different approaches is similar for all $U/t$, however the key aspect is the amplitude of the $m_\alpha$ calculated using DMFT is significantly less than to the one obtained from  MF approach. This is due to the many-body corrections included in DMFT, not accounted by the mean-field calculations.
Interestingly, and despite such differences, it is observed that the flat band electron gas does not
fully screen the local moment.

The hybridization between the electrons at non-interacting sites and the impurity site leads to the formation of the singlet between the spins of electrons at different sites. In contrast, the RKKY interaction makes the impurity spins interacting with each other via conduction electrons and thus tends to stabilize the magnetic ordering. 
This can be shown by the finite hybridization along with the local magnetic order at $U/t$ (Fig.~\ref{fig:hyb_gap}).  Effective hybridization decreases monotonically both at impurity sites and site next to the impurity such that $\Delta_I < \Delta_{A/B}$ for all $U/t$ consistent with the magnetic order at corresponding site. Effective hybridization at the second neighbor site decreases initially and then  increases further, congruous with magnetic order.

\subsection{Finite temperature calculations}
\label{sec:temp}
\begin{figure}
\includegraphics[width=\columnwidth]{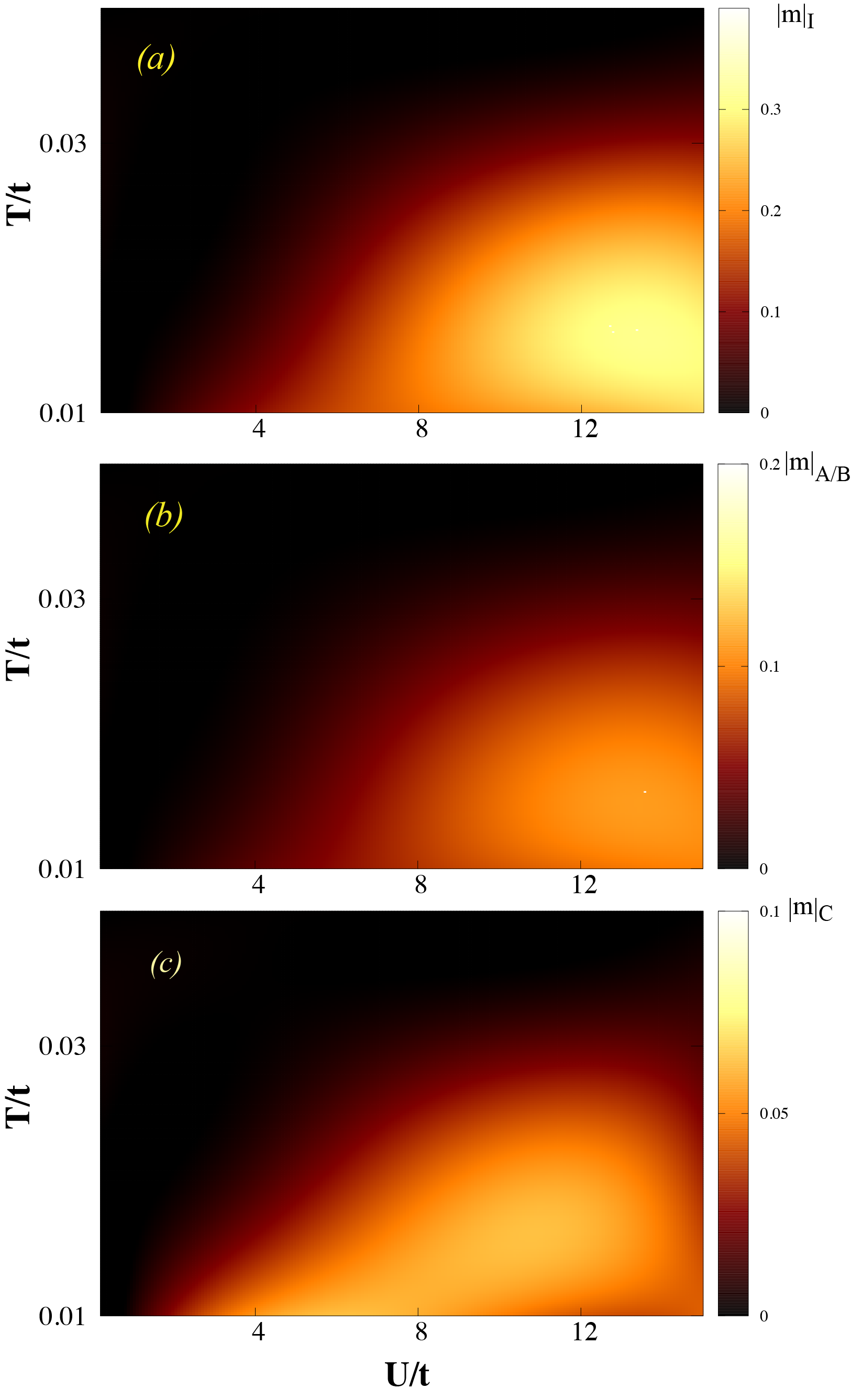}
\caption{Temperature \textit{vs} interaction strength phase diagram of the inhomogeneous Hubbard model showing local magnetization at (a) impurity site, $m_{\text{I}}$, (b) at site next to the impurity, $m_{\text{A/B}}$ and (c) second neighbor site to the impurity site, $m_{\text{C}}$.}
\label{fig5}
\end{figure}
We now move on to consider the effects of finite temperature in the dynamical
mean-field calculations. First, it is worth to note that
the current model will not show long range magnetic order
due to its two-dimensional nature. Dynamical mean-field
theory however does not capture long-range fluctuations,
and therefore does not account for Mermin-Wagner theorem\cite{PhysRevLett.17.1133,PhysRev.158.383}.
Within this framework, the existence of a finite-magnetization
in a DMFT calculation must be rationalized as a signature of
strong magnetic correlations, as addressed in other systems\cite{PhysRevB.95.235109}.
In this context, while the true long range ordering temperature is at $T=0$,
correlation lengths start to grow at a specific temperature scale
$T^*>0$, and is the temperature scale captured by DMFT\cite{PhysRevB.95.235109}.

\begin{figure}
\includegraphics[width=\columnwidth]{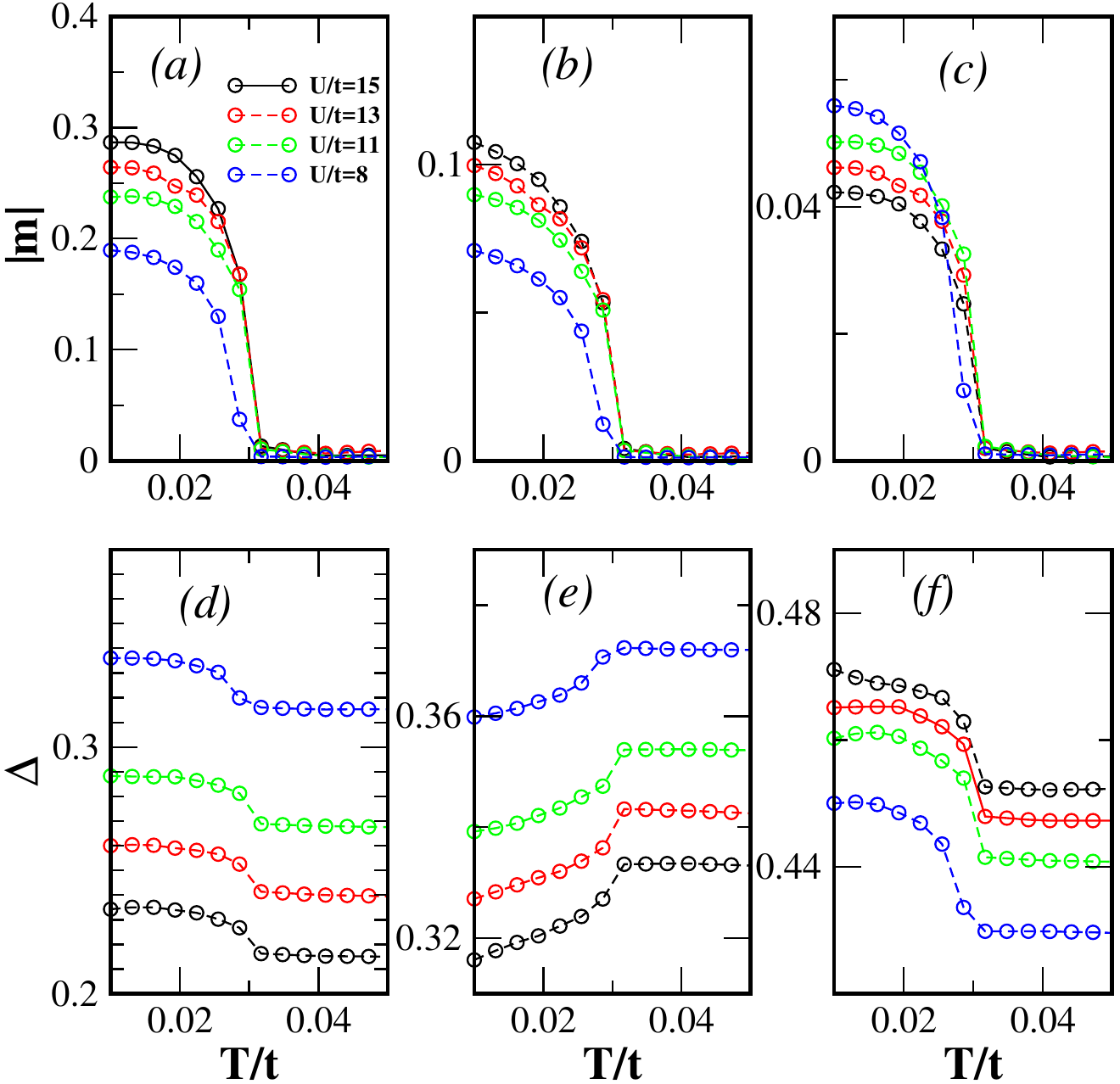}
\caption{Local magnetic order (a, b, c) and the corresponding effective hybridization (c, d, e) for the impurity site (a,c), site next to impurity site (b,d) and second neighbor site (c,e) of the impurity with varying temperature.}
\label{fig6} 
\end{figure}

In Fig.~\ref{fig5}, we show the region in the T-U plane where the magnetic order, $m_\alpha$, is finite.
The amplitude of $m_\alpha$, at a given $U/t$, is site dependent. Local magnetization at the impurity site and at the site next to the impurity monotonically increases and then saturates with increasing interaction strength with a nonmagnetic to magnetic transition at $T_c$. However, the magnetic behavior at the second neighbor site gets peaked at $U_p/t$ for a given $T$. $U_p$ increases monotonically with the $T$.
We show the spatially resolved magnetic order $m_\alpha$  the corresponding effective  hybridization evaluated using CTQMC+CDMFT with varying temperature at different interaction strength $U/t$ in the upper panels and lower panels of the Fig.~\ref{fig6} respectively. As the temperature is increased, the magnetic order for all the sublattices sharply disappears at $T_c$. The evaluation of the magnetic order at the second neighbor site is decrease with increasing  value of $U$ for a given $T$, opposite to the impurity site and the site next to the impurity. Effective hybridization changes with the transition temperature, with a cusp at $T_c$. 
It is important to note that the effective hybridization is finite in both magnetic and nonmagnetic regions, and the behavior of $\Delta$ is consistent with the corresponding local magnetic order.

\section{Tensor-network approach to the one-dimensional flat band Kondo lattice model}
\label{sec:1d}

\label{oned:model}
\begin{figure}
\center
\includegraphics[width=\columnwidth]{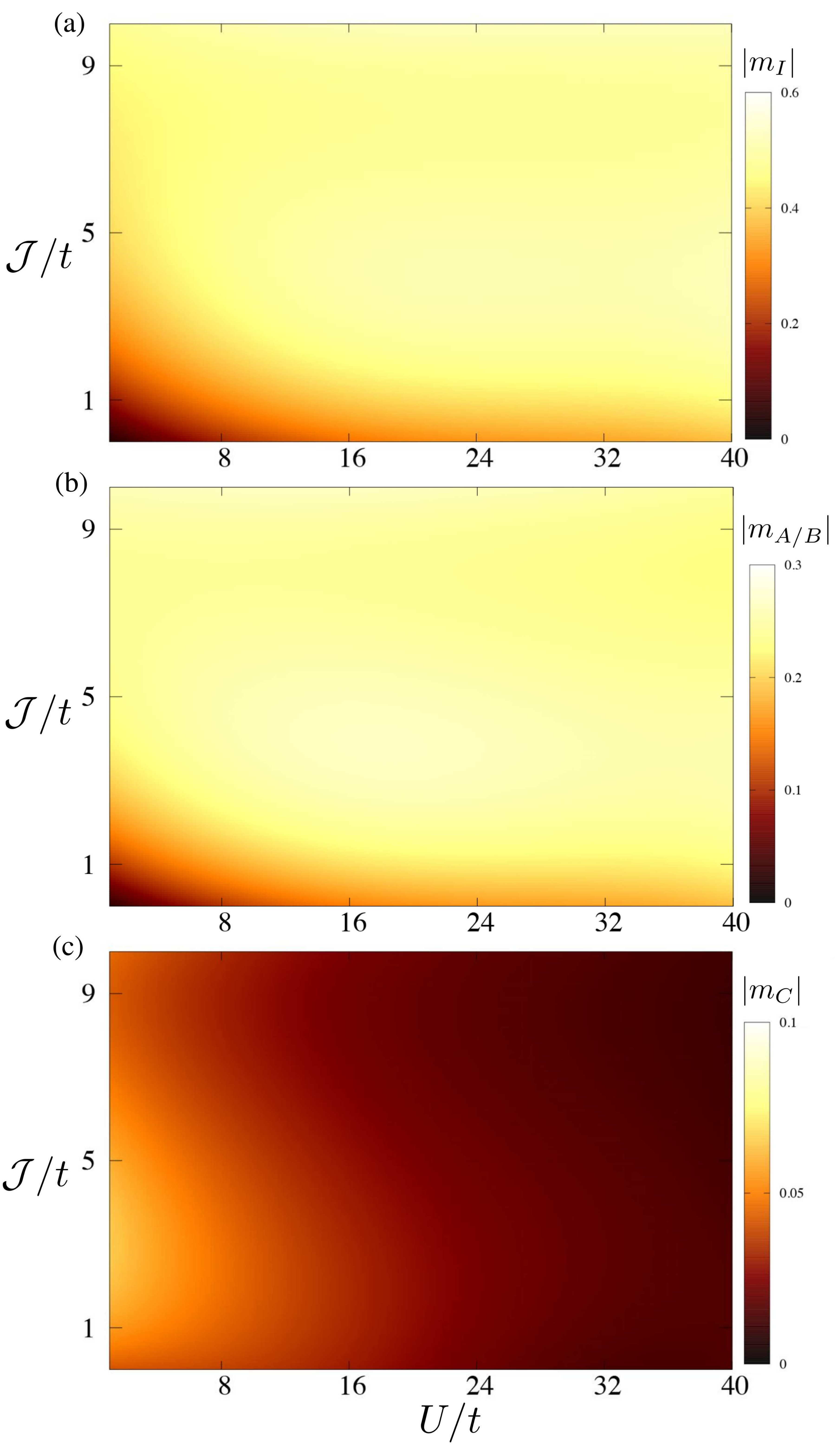}
\caption{Local  magnetic order on (a) impurity site I, (b) flat-band sites A and B, and (c) site C v.s. Hubbard U and local exchange field $\B/t$ for the equivalent one dimensional model.}
\label{fig7:dmrg}
\end{figure}

We now move on to consider a one dimensional flat band Kondo lattice model,
which can be rationalized as a one-dimensional version of the model outlined above.
First, it is worth emphasizing that for a one-dimensional
quantum many-body problem, the existence of strong quantum many-body fluctuations
would prevent the system to order even in the presence of strong magnetic
correlations. To accommodate this restriction, we will
generalize the flat band Kondo lattice model, including local symmetry breaking terms,
allowing for a selective quench of the quantum fluctuations of the system.

Let us now comment on the specific model we will address.
On the computational side, the interacting model can be effectively solved using tensor network
methods\cite{PhysRevLett.69.2863,2020arXiv200714822F,ITensor,dmrgpy} due to its one-dimensional nature. 
We will include a non-interacting hopping term in a geometry shown in Fig. \ref{fig:fig1}c,
and whose flat band structure is shown in Fig. \ref{fig:fig1}f.
Local Hubbard interactions are included in the site $I$ as depicted in  Fig. \ref{fig:fig1}d.
On top of the interacting term,
we will also include a local exchange field on an impurity site, leading to a full
Hamiltonian of the form

\begin{equation}
\mathcal{H}=t\sum_{\langle i,j\rangle}c^\dag_ic_j+
U\sum_{i\in I}
\left (n_{i\uparrow}-\frac{1}{2} \right )
\left (n_{i\downarrow}-\frac{1}{2} \right )
+ 
\B \sum_I S^z_{i}
\end{equation}

\begin{figure}
\center
\includegraphics[width=\columnwidth]{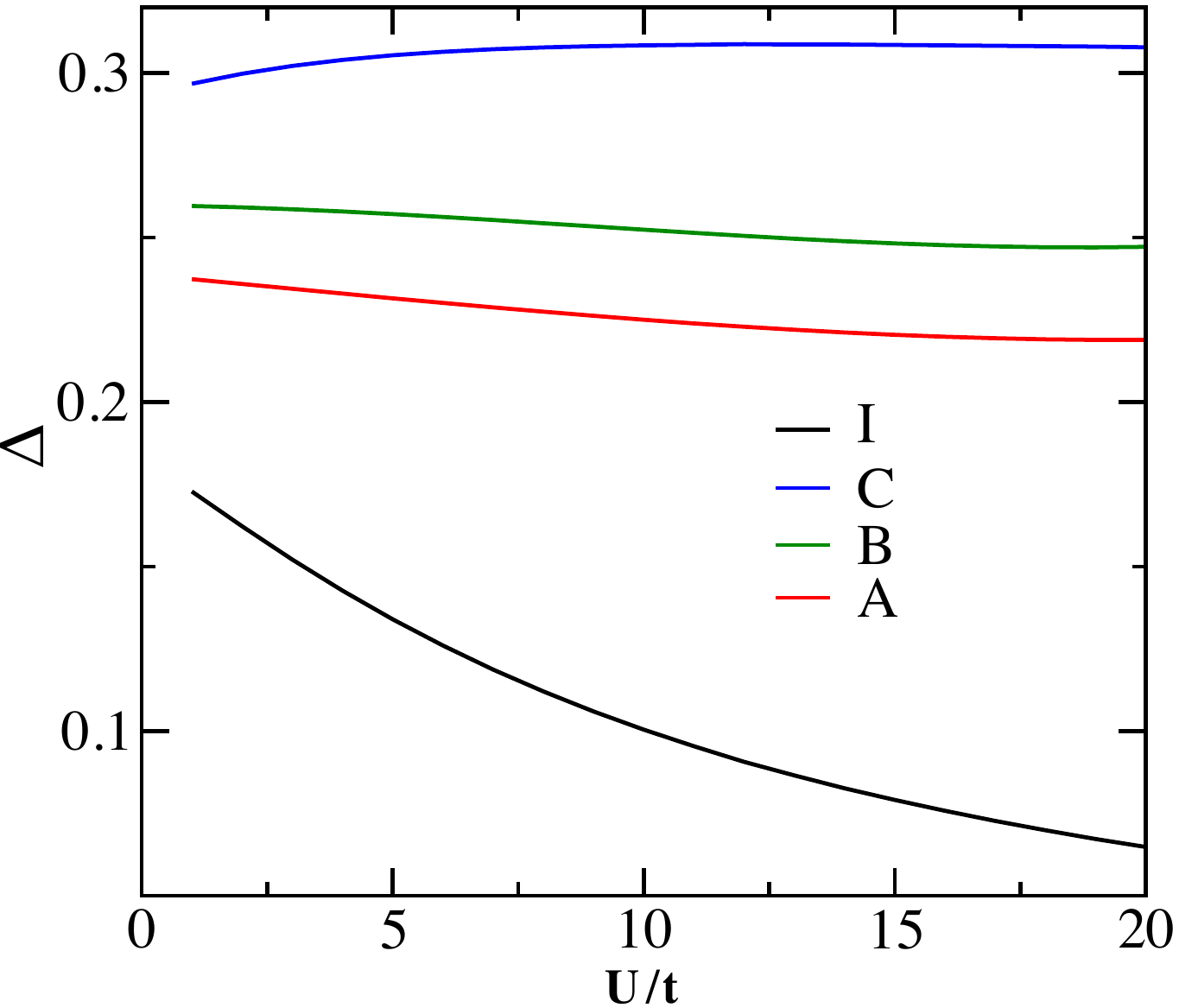}
\caption{Effective hybridization at different sublattice sites in the one-dimensional flat band Kondo lattice model with $\B=5t$ for varying $U/t$.}
\label{fig8}
\end{figure}

Increasing the local exchange
$\B$ can lead to the emergence of the quasi-flat band at the site next to the
impurity site, even at $U=0$.
In a mean-field language, $\B$ could be rationalized as an effective local magnetization, and therefore
$\B$ allows to effectively interpolate between a mean-field and purely many-body limit.
 In particular, when $\B\to\infty$, the impurity site is fully magnetized, and we are in pure mean-field limit, whereas when $\B=0$, the magnetic instability solely originates from $U$ and we are in a pure many-body limit.
To be concrete, we consider a chain of 40 sites (i.e., 20 lattice constants long) and compute averaged magnetization on the impurity sites $I$, the flat-band sites $A$ and $B$, and the next the nearest site C with varying Hubbard $U$ and the field $\B$ of impurity sites. We find that the local exchange $\B$ destroys quantum fluctuations and promotes a quasi-long range order at the impurity site and sites $A/B$. 
As $\B$ becomes smaller, larger $U$ is required to produce the magnetic order.

 The above behavior is consistent with the hybridization
 effects in the two-dimensional
 flat band Kondo lattice addressed previously with dynamical mean-field theory. 
 In particular, we show the effective hybridization varying with $U$ at fixed field $\B=5t$ in Fig.\ref{fig8}. We find that the effective hybridization at all sublattice sites has the same behavior as in the 2D case in Fig.\ref{fig:hyb_gap}. Due to the finite exchange field $\B$, the hybridization at all sites are suppressed, with the impurity site having the largest suppression. The previous results show that, while the one-dimensional model
 is formally different from its two-dimensional counterpart, magnetic correlations between Kondo sites
 and flat band sites show a similar phenomenology.

\section{Conclusion}
\label{sec:con}
Kondo lattice problems represent one of the paradigmatic
many-body problems, hosting a variety
of intricate phenomena due to the competitions of
Kondo screening and magnetic ordering. 
Here, we have addressed the physics
of a Kondo lattice problem, in which a conventional dispersive electron gas is
replaced by a flat band electronic state. In particular, we considered 
flat band Kondo lattice models in both two and one dimensions, which we solved using many-body dynamical mean-field theory and tensor networks, respectively. 
We have demonstrated the emergence of a
locally ordered state, showing how the formation of the Kondo cloud is overcome by
the magnetic correlation between impurities.
Interestingly, the underscreened phase
observed in our many-body results can be captured
via a conventional symmetry broken mean-field method, highlighting how the
main features of the system are qualitatively reproduced.
Our treatment focused purely on the half-filled case,
yet doping of the Kondo lattice is expected to lead to
a more complex interplay of the exchange and Kondo coupling scales,
therefore leading to substantially more complex phase diagrams.
Our results exemplify the fine interplay between magnetic
ordering and many-body screening, putting forward flat band Kondo
lattice models as a powerful platform to explore
exotic emergent quantum many-body states.

\textbf{Acknowledgments:}
We acknowledge the computational resources provided by
the Aalto Science-IT project. J. L. L. acknowledges
financial support from the Academy of Finland Projects No.
331342 and No. 336243, and the Jane and Aatos
Erkko Foundation. We thank A. Ramires, P. Törmä,
P. Liljeroth and M. Aapro for useful discussions.

\bibliography{biblio}

\begin{thebibliography}{71}%
\makeatletter
\providecommand \@ifxundefined [1]{%
 \@ifx{#1\undefined}
}%
\providecommand \@ifnum [1]{%
 \ifnum #1\expandafter \@firstoftwo
 \else \expandafter \@secondoftwo
 \fi
}%
\providecommand \@ifx [1]{%
 \ifx #1\expandafter \@firstoftwo
 \else \expandafter \@secondoftwo
 \fi
}%
\providecommand \natexlab [1]{#1}%
\providecommand \enquote  [1]{``#1''}%
\providecommand \bibnamefont  [1]{#1}%
\providecommand \bibfnamefont [1]{#1}%
\providecommand \citenamefont [1]{#1}%
\providecommand \href@noop [0]{\@secondoftwo}%
\providecommand \href [0]{\begingroup \@sanitize@url \@href}%
\providecommand \@href[1]{\@@startlink{#1}\@@href}%
\providecommand \@@href[1]{\endgroup#1\@@endlink}%
\providecommand \@sanitize@url [0]{\catcode `\\12\catcode `\$12\catcode
  `\&12\catcode `\#12\catcode `\^12\catcode `\_12\catcode `\%12\relax}%
\providecommand \@@startlink[1]{}%
\providecommand \@@endlink[0]{}%
\providecommand \url  [0]{\begingroup\@sanitize@url \@url }%
\providecommand \@url [1]{\endgroup\@href {#1}{\urlprefix }}%
\providecommand \urlprefix  [0]{URL }%
\providecommand \Eprint [0]{\href }%
\providecommand \doibase [0]{https://doi.org/}%
\providecommand \selectlanguage [0]{\@gobble}%
\providecommand \bibinfo  [0]{\@secondoftwo}%
\providecommand \bibfield  [0]{\@secondoftwo}%
\providecommand \translation [1]{[#1]}%
\providecommand \BibitemOpen [0]{}%
\providecommand \bibitemStop [0]{}%
\providecommand \bibitemNoStop [0]{.\EOS\space}%
\providecommand \EOS [0]{\spacefactor3000\relax}%
\providecommand \BibitemShut  [1]{\csname bibitem#1\endcsname}%
\let\auto@bib@innerbib\@empty
\bibitem [{\citenamefont {Ruostekoski}(2009)}]{PhysRevLett.103.080406}%
  \BibitemOpen
  \bibfield  {author} {\bibinfo {author} {\bibfnamefont {J.}~\bibnamefont
  {Ruostekoski}},\ }\bibfield  {title} {\bibinfo {title} {Optical kagome
  lattice for ultracold atoms with nearest neighbor interactions},\ }\href
  {https://doi.org/10.1103/PhysRevLett.103.080406} {\bibfield  {journal}
  {\bibinfo  {journal} {Phys. Rev. Lett.}\ }\textbf {\bibinfo {volume} {103}},\
  \bibinfo {pages} {080406} (\bibinfo {year} {2009})}\BibitemShut {NoStop}%
\bibitem [{\citenamefont {Iglovikov}\ \emph {et~al.}(2014)\citenamefont
  {Iglovikov}, \citenamefont {H\'ebert}, \citenamefont {Gr\'emaud},
  \citenamefont {Batrouni},\ and\ \citenamefont
  {Scalettar}}]{PhysRevB.90.094506}%
  \BibitemOpen
  \bibfield  {author} {\bibinfo {author} {\bibfnamefont {V.~I.}\ \bibnamefont
  {Iglovikov}}, \bibinfo {author} {\bibfnamefont {F.}~\bibnamefont {H\'ebert}},
  \bibinfo {author} {\bibfnamefont {B.}~\bibnamefont {Gr\'emaud}}, \bibinfo
  {author} {\bibfnamefont {G.~G.}\ \bibnamefont {Batrouni}},\ and\ \bibinfo
  {author} {\bibfnamefont {R.~T.}\ \bibnamefont {Scalettar}},\ }\bibfield
  {title} {\bibinfo {title} {Superconducting transitions in flat-band
  systems},\ }\href {https://doi.org/10.1103/PhysRevB.90.094506} {\bibfield
  {journal} {\bibinfo  {journal} {Phys. Rev. B}\ }\textbf {\bibinfo {volume}
  {90}},\ \bibinfo {pages} {094506} (\bibinfo {year} {2014})}\BibitemShut
  {NoStop}%
\bibitem [{\citenamefont {Kopnin}\ \emph
  {et~al.}(2011{\natexlab{a}})\citenamefont {Kopnin}, \citenamefont
  {Heikkil\"a},\ and\ \citenamefont {Volovik}}]{PhysRevB.83.220503}%
  \BibitemOpen
  \bibfield  {author} {\bibinfo {author} {\bibfnamefont {N.~B.}\ \bibnamefont
  {Kopnin}}, \bibinfo {author} {\bibfnamefont {T.~T.}\ \bibnamefont
  {Heikkil\"a}},\ and\ \bibinfo {author} {\bibfnamefont {G.~E.}\ \bibnamefont
  {Volovik}},\ }\bibfield  {title} {\bibinfo {title} {High-temperature surface
  superconductivity in topological flat-band systems},\ }\href
  {https://doi.org/10.1103/PhysRevB.83.220503} {\bibfield  {journal} {\bibinfo
  {journal} {Phys. Rev. B}\ }\textbf {\bibinfo {volume} {83}},\ \bibinfo
  {pages} {220503} (\bibinfo {year} {2011}{\natexlab{a}})}\BibitemShut
  {NoStop}%
\bibitem [{\citenamefont {Tovmasyan}\ \emph {et~al.}(2016)\citenamefont
  {Tovmasyan}, \citenamefont {Peotta}, \citenamefont {T\"orm\"a},\ and\
  \citenamefont {Huber}}]{PhysRevB.94.245149}%
  \BibitemOpen
  \bibfield  {author} {\bibinfo {author} {\bibfnamefont {M.}~\bibnamefont
  {Tovmasyan}}, \bibinfo {author} {\bibfnamefont {S.}~\bibnamefont {Peotta}},
  \bibinfo {author} {\bibfnamefont {P.}~\bibnamefont {T\"orm\"a}},\ and\
  \bibinfo {author} {\bibfnamefont {S.~D.}\ \bibnamefont {Huber}},\ }\bibfield
  {title} {\bibinfo {title} {Effective theory and emergent $\text{SU}(2)$
  symmetry in the flat bands of attractive hubbard models},\ }\href
  {https://doi.org/10.1103/PhysRevB.94.245149} {\bibfield  {journal} {\bibinfo
  {journal} {Phys. Rev. B}\ }\textbf {\bibinfo {volume} {94}},\ \bibinfo
  {pages} {245149} (\bibinfo {year} {2016})}\BibitemShut {NoStop}%
\bibitem [{\citenamefont {Hofmann}\ \emph {et~al.}(2020)\citenamefont
  {Hofmann}, \citenamefont {Berg},\ and\ \citenamefont
  {Chowdhury}}]{PhysRevB.102.201112}%
  \BibitemOpen
  \bibfield  {author} {\bibinfo {author} {\bibfnamefont {J.~S.}\ \bibnamefont
  {Hofmann}}, \bibinfo {author} {\bibfnamefont {E.}~\bibnamefont {Berg}},\ and\
  \bibinfo {author} {\bibfnamefont {D.}~\bibnamefont {Chowdhury}},\ }\bibfield
  {title} {\bibinfo {title} {Superconductivity, pseudogap, and phase separation
  in topological flat bands},\ }\href
  {https://doi.org/10.1103/PhysRevB.102.201112} {\bibfield  {journal} {\bibinfo
   {journal} {Phys. Rev. B}\ }\textbf {\bibinfo {volume} {102}},\ \bibinfo
  {pages} {201112} (\bibinfo {year} {2020})}\BibitemShut {NoStop}%
\bibitem [{\citenamefont {Peri}\ \emph {et~al.}(2021)\citenamefont {Peri},
  \citenamefont {Song}, \citenamefont {Bernevig},\ and\ \citenamefont
  {Huber}}]{PhysRevLett.126.027002}%
  \BibitemOpen
  \bibfield  {author} {\bibinfo {author} {\bibfnamefont {V.}~\bibnamefont
  {Peri}}, \bibinfo {author} {\bibfnamefont {Z.-D.}\ \bibnamefont {Song}},
  \bibinfo {author} {\bibfnamefont {B.~A.}\ \bibnamefont {Bernevig}},\ and\
  \bibinfo {author} {\bibfnamefont {S.~D.}\ \bibnamefont {Huber}},\ }\bibfield
  {title} {\bibinfo {title} {Fragile topology and flat-band superconductivity
  in the strong-coupling regime},\ }\href
  {https://doi.org/10.1103/PhysRevLett.126.027002} {\bibfield  {journal}
  {\bibinfo  {journal} {Phys. Rev. Lett.}\ }\textbf {\bibinfo {volume} {126}},\
  \bibinfo {pages} {027002} (\bibinfo {year} {2021})}\BibitemShut {NoStop}%
\bibitem [{\citenamefont {Drost}\ \emph {et~al.}(2017)\citenamefont {Drost},
  \citenamefont {Ojanen}, \citenamefont {Harju},\ and\ \citenamefont
  {Liljeroth}}]{Drost2017}%
  \BibitemOpen
  \bibfield  {author} {\bibinfo {author} {\bibfnamefont {R.}~\bibnamefont
  {Drost}}, \bibinfo {author} {\bibfnamefont {T.}~\bibnamefont {Ojanen}},
  \bibinfo {author} {\bibfnamefont {A.}~\bibnamefont {Harju}},\ and\ \bibinfo
  {author} {\bibfnamefont {P.}~\bibnamefont {Liljeroth}},\ }\bibfield  {title}
  {\bibinfo {title} {Topological states in engineered atomic lattices},\ }\href
  {https://doi.org/10.1038/nphys4080} {\bibfield  {journal} {\bibinfo
  {journal} {Nature Physics}\ }\textbf {\bibinfo {volume} {13}},\ \bibinfo
  {pages} {668} (\bibinfo {year} {2017})}\BibitemShut {NoStop}%
\bibitem [{\citenamefont {Slot}\ \emph {et~al.}(2017)\citenamefont {Slot},
  \citenamefont {Gardenier}, \citenamefont {Jacobse}, \citenamefont {van
  Miert}, \citenamefont {Kempkes}, \citenamefont {Zevenhuizen}, \citenamefont
  {Smith}, \citenamefont {Vanmaekelbergh},\ and\ \citenamefont
  {Swart}}]{Slot2017}%
  \BibitemOpen
  \bibfield  {author} {\bibinfo {author} {\bibfnamefont {M.~R.}\ \bibnamefont
  {Slot}}, \bibinfo {author} {\bibfnamefont {T.~S.}\ \bibnamefont {Gardenier}},
  \bibinfo {author} {\bibfnamefont {P.~H.}\ \bibnamefont {Jacobse}}, \bibinfo
  {author} {\bibfnamefont {G.~C.~P.}\ \bibnamefont {van Miert}}, \bibinfo
  {author} {\bibfnamefont {S.~N.}\ \bibnamefont {Kempkes}}, \bibinfo {author}
  {\bibfnamefont {S.~J.~M.}\ \bibnamefont {Zevenhuizen}}, \bibinfo {author}
  {\bibfnamefont {C.~M.}\ \bibnamefont {Smith}}, \bibinfo {author}
  {\bibfnamefont {D.}~\bibnamefont {Vanmaekelbergh}},\ and\ \bibinfo {author}
  {\bibfnamefont {I.}~\bibnamefont {Swart}},\ }\bibfield  {title} {\bibinfo
  {title} {Experimental realization and characterization of an electronic lieb
  lattice},\ }\href {https://doi.org/10.1038/nphys4105} {\bibfield  {journal}
  {\bibinfo  {journal} {Nature Physics}\ }\textbf {\bibinfo {volume} {13}},\
  \bibinfo {pages} {672} (\bibinfo {year} {2017})}\BibitemShut {NoStop}%
\bibitem [{\citenamefont {Huda}\ \emph {et~al.}(2020)\citenamefont {Huda},
  \citenamefont {Kezilebieke},\ and\ \citenamefont
  {Liljeroth}}]{PhysRevResearch.2.043426}%
  \BibitemOpen
  \bibfield  {author} {\bibinfo {author} {\bibfnamefont {M.~N.}\ \bibnamefont
  {Huda}}, \bibinfo {author} {\bibfnamefont {S.}~\bibnamefont {Kezilebieke}},\
  and\ \bibinfo {author} {\bibfnamefont {P.}~\bibnamefont {Liljeroth}},\
  }\bibfield  {title} {\bibinfo {title} {Designer flat bands in
  quasi-one-dimensional atomic lattices},\ }\href
  {https://doi.org/10.1103/PhysRevResearch.2.043426} {\bibfield  {journal}
  {\bibinfo  {journal} {Phys. Rev. Research}\ }\textbf {\bibinfo {volume}
  {2}},\ \bibinfo {pages} {043426} (\bibinfo {year} {2020})}\BibitemShut
  {NoStop}%
\bibitem [{\citenamefont {Khajetoorians}\ \emph {et~al.}(2019)\citenamefont
  {Khajetoorians}, \citenamefont {Wegner}, \citenamefont {Otte},\ and\
  \citenamefont {Swart}}]{Khajetoorians2019}%
  \BibitemOpen
  \bibfield  {author} {\bibinfo {author} {\bibfnamefont {A.~A.}\ \bibnamefont
  {Khajetoorians}}, \bibinfo {author} {\bibfnamefont {D.}~\bibnamefont
  {Wegner}}, \bibinfo {author} {\bibfnamefont {A.~F.}\ \bibnamefont {Otte}},\
  and\ \bibinfo {author} {\bibfnamefont {I.}~\bibnamefont {Swart}},\ }\bibfield
   {title} {\bibinfo {title} {Creating designer quantum states of matter
  atom-by-atom},\ }\href {https://doi.org/10.1038/s42254-019-0108-5} {\bibfield
   {journal} {\bibinfo  {journal} {Nature Reviews Physics}\ }\textbf {\bibinfo
  {volume} {1}},\ \bibinfo {pages} {703} (\bibinfo {year} {2019})}\BibitemShut
  {NoStop}%
\bibitem [{\citenamefont {Jo}\ \emph {et~al.}(2012)\citenamefont {Jo},
  \citenamefont {Guzman}, \citenamefont {Thomas}, \citenamefont {Hosur},
  \citenamefont {Vishwanath},\ and\ \citenamefont
  {Stamper-Kurn}}]{PhysRevLett.108.045305}%
  \BibitemOpen
  \bibfield  {author} {\bibinfo {author} {\bibfnamefont {G.-B.}\ \bibnamefont
  {Jo}}, \bibinfo {author} {\bibfnamefont {J.}~\bibnamefont {Guzman}}, \bibinfo
  {author} {\bibfnamefont {C.~K.}\ \bibnamefont {Thomas}}, \bibinfo {author}
  {\bibfnamefont {P.}~\bibnamefont {Hosur}}, \bibinfo {author} {\bibfnamefont
  {A.}~\bibnamefont {Vishwanath}},\ and\ \bibinfo {author} {\bibfnamefont
  {D.~M.}\ \bibnamefont {Stamper-Kurn}},\ }\bibfield  {title} {\bibinfo {title}
  {Ultracold atoms in a tunable optical kagome lattice},\ }\href
  {https://doi.org/10.1103/PhysRevLett.108.045305} {\bibfield  {journal}
  {\bibinfo  {journal} {Phys. Rev. Lett.}\ }\textbf {\bibinfo {volume} {108}},\
  \bibinfo {pages} {045305} (\bibinfo {year} {2012})}\BibitemShut {NoStop}%
\bibitem [{\citenamefont {Taie}\ \emph {et~al.}(2015)\citenamefont {Taie},
  \citenamefont {Ozawa}, \citenamefont {Ichinose}, \citenamefont {Nishio},
  \citenamefont {Nakajima},\ and\ \citenamefont {Takahashi}}]{Taie2015}%
  \BibitemOpen
  \bibfield  {author} {\bibinfo {author} {\bibfnamefont {S.}~\bibnamefont
  {Taie}}, \bibinfo {author} {\bibfnamefont {H.}~\bibnamefont {Ozawa}},
  \bibinfo {author} {\bibfnamefont {T.}~\bibnamefont {Ichinose}}, \bibinfo
  {author} {\bibfnamefont {T.}~\bibnamefont {Nishio}}, \bibinfo {author}
  {\bibfnamefont {S.}~\bibnamefont {Nakajima}},\ and\ \bibinfo {author}
  {\bibfnamefont {Y.}~\bibnamefont {Takahashi}},\ }\bibfield  {title} {\bibinfo
  {title} {Coherent driving and freezing of bosonic matter wave in an optical
  lieb lattice},\ }\href {https://doi.org/10.1126/sciadv.1500854} {\bibfield
  {journal} {\bibinfo  {journal} {Science Advances}\ }\textbf {\bibinfo
  {volume} {1}},\ \bibinfo {pages} {e1500854} (\bibinfo {year}
  {2015})}\BibitemShut {NoStop}%
\bibitem [{\citenamefont {Bloch}\ \emph {et~al.}(2012)\citenamefont {Bloch},
  \citenamefont {Dalibard},\ and\ \citenamefont
  {Nascimb{\`{e}}ne}}]{Bloch2012}%
  \BibitemOpen
  \bibfield  {author} {\bibinfo {author} {\bibfnamefont {I.}~\bibnamefont
  {Bloch}}, \bibinfo {author} {\bibfnamefont {J.}~\bibnamefont {Dalibard}},\
  and\ \bibinfo {author} {\bibfnamefont {S.}~\bibnamefont {Nascimb{\`{e}}ne}},\
  }\bibfield  {title} {\bibinfo {title} {Quantum simulations with ultracold
  quantum gases},\ }\href {https://doi.org/10.1038/nphys2259} {\bibfield
  {journal} {\bibinfo  {journal} {Nature Physics}\ }\textbf {\bibinfo {volume}
  {8}},\ \bibinfo {pages} {267} (\bibinfo {year} {2012})}\BibitemShut {NoStop}%
\bibitem [{\citenamefont {Su\'arez~Morell}\ \emph {et~al.}(2010)\citenamefont
  {Su\'arez~Morell}, \citenamefont {Correa}, \citenamefont {Vargas},
  \citenamefont {Pacheco},\ and\ \citenamefont
  {Barticevic}}]{PhysRevB.82.121407}%
  \BibitemOpen
  \bibfield  {author} {\bibinfo {author} {\bibfnamefont {E.}~\bibnamefont
  {Su\'arez~Morell}}, \bibinfo {author} {\bibfnamefont {J.~D.}\ \bibnamefont
  {Correa}}, \bibinfo {author} {\bibfnamefont {P.}~\bibnamefont {Vargas}},
  \bibinfo {author} {\bibfnamefont {M.}~\bibnamefont {Pacheco}},\ and\ \bibinfo
  {author} {\bibfnamefont {Z.}~\bibnamefont {Barticevic}},\ }\bibfield  {title}
  {\bibinfo {title} {Flat bands in slightly twisted bilayer graphene:
  Tight-binding calculations},\ }\href
  {https://doi.org/10.1103/PhysRevB.82.121407} {\bibfield  {journal} {\bibinfo
  {journal} {Phys. Rev. B}\ }\textbf {\bibinfo {volume} {82}},\ \bibinfo
  {pages} {121407} (\bibinfo {year} {2010})}\BibitemShut {NoStop}%
\bibitem [{\citenamefont {Cao}\ \emph {et~al.}(2018{\natexlab{a}})\citenamefont
  {Cao}, \citenamefont {Fatemi}, \citenamefont {Fang}, \citenamefont
  {Watanabe}, \citenamefont {Taniguchi}, \citenamefont {Kaxiras},\ and\
  \citenamefont {Jarillo-Herrero}}]{Cao2018super}%
  \BibitemOpen
  \bibfield  {author} {\bibinfo {author} {\bibfnamefont {Y.}~\bibnamefont
  {Cao}}, \bibinfo {author} {\bibfnamefont {V.}~\bibnamefont {Fatemi}},
  \bibinfo {author} {\bibfnamefont {S.}~\bibnamefont {Fang}}, \bibinfo {author}
  {\bibfnamefont {K.}~\bibnamefont {Watanabe}}, \bibinfo {author}
  {\bibfnamefont {T.}~\bibnamefont {Taniguchi}}, \bibinfo {author}
  {\bibfnamefont {E.}~\bibnamefont {Kaxiras}},\ and\ \bibinfo {author}
  {\bibfnamefont {P.}~\bibnamefont {Jarillo-Herrero}},\ }\bibfield  {title}
  {\bibinfo {title} {Unconventional superconductivity in magic-angle graphene
  superlattices},\ }\href {https://doi.org/10.1038/nature26160} {\bibfield
  {journal} {\bibinfo  {journal} {Nature}\ }\textbf {\bibinfo {volume} {556}},\
  \bibinfo {pages} {43} (\bibinfo {year} {2018}{\natexlab{a}})}\BibitemShut
  {NoStop}%
\bibitem [{\citenamefont {Andrei}\ \emph {et~al.}(2021)\citenamefont {Andrei},
  \citenamefont {Efetov}, \citenamefont {Jarillo-Herrero}, \citenamefont
  {MacDonald}, \citenamefont {Mak}, \citenamefont {Senthil}, \citenamefont
  {Tutuc}, \citenamefont {Yazdani},\ and\ \citenamefont {Young}}]{Andrei2021}%
  \BibitemOpen
  \bibfield  {author} {\bibinfo {author} {\bibfnamefont {E.~Y.}\ \bibnamefont
  {Andrei}}, \bibinfo {author} {\bibfnamefont {D.~K.}\ \bibnamefont {Efetov}},
  \bibinfo {author} {\bibfnamefont {P.}~\bibnamefont {Jarillo-Herrero}},
  \bibinfo {author} {\bibfnamefont {A.~H.}\ \bibnamefont {MacDonald}}, \bibinfo
  {author} {\bibfnamefont {K.~F.}\ \bibnamefont {Mak}}, \bibinfo {author}
  {\bibfnamefont {T.}~\bibnamefont {Senthil}}, \bibinfo {author} {\bibfnamefont
  {E.}~\bibnamefont {Tutuc}}, \bibinfo {author} {\bibfnamefont
  {A.}~\bibnamefont {Yazdani}},\ and\ \bibinfo {author} {\bibfnamefont {A.~F.}\
  \bibnamefont {Young}},\ }\bibfield  {title} {\bibinfo {title} {The marvels of
  moir{\'{e}} materials},\ }\href {https://doi.org/10.1038/s41578-021-00284-1}
  {\bibfield  {journal} {\bibinfo  {journal} {Nature Reviews Materials}\
  }\textbf {\bibinfo {volume} {6}},\ \bibinfo {pages} {201} (\bibinfo {year}
  {2021})}\BibitemShut {NoStop}%
\bibitem [{\citenamefont {Leykam}\ \emph {et~al.}(2018)\citenamefont {Leykam},
  \citenamefont {Andreanov},\ and\ \citenamefont {Flach}}]{Leykam2018}%
  \BibitemOpen
  \bibfield  {author} {\bibinfo {author} {\bibfnamefont {D.}~\bibnamefont
  {Leykam}}, \bibinfo {author} {\bibfnamefont {A.}~\bibnamefont {Andreanov}},\
  and\ \bibinfo {author} {\bibfnamefont {S.}~\bibnamefont {Flach}},\ }\bibfield
   {title} {\bibinfo {title} {Artificial flat band systems: from lattice models
  to experiments},\ }\href {https://doi.org/10.1080/23746149.2018.1473052}
  {\bibfield  {journal} {\bibinfo  {journal} {Advances in Physics: X}\ }\textbf
  {\bibinfo {volume} {3}},\ \bibinfo {pages} {1473052} (\bibinfo {year}
  {2018})},\ \Eprint
  {https://arxiv.org/abs/https://doi.org/10.1080/23746149.2018.1473052}
  {https://doi.org/10.1080/23746149.2018.1473052} \BibitemShut {NoStop}%
\bibitem [{\citenamefont {Kopnin}\ \emph
  {et~al.}(2011{\natexlab{b}})\citenamefont {Kopnin}, \citenamefont
  {Heikkil\"a},\ and\ \citenamefont {Volovik}}]{Kopnin:2011}%
  \BibitemOpen
  \bibfield  {author} {\bibinfo {author} {\bibfnamefont {N.~B.}\ \bibnamefont
  {Kopnin}}, \bibinfo {author} {\bibfnamefont {T.~T.}\ \bibnamefont
  {Heikkil\"a}},\ and\ \bibinfo {author} {\bibfnamefont {G.~E.}\ \bibnamefont
  {Volovik}},\ }\bibfield  {title} {\bibinfo {title} {{High-temperature surface
  superconductivity in topological flat-band systems}},\ }\href
  {http://link.aps.org/doi/10.1103/PhysRevB.83.220503} {\bibfield  {journal}
  {\bibinfo  {journal} {Phys. Rev. B}\ }\textbf {\bibinfo {volume} {83}},\
  \bibinfo {pages} {220503} (\bibinfo {year} {2011}{\natexlab{b}})}\BibitemShut
  {NoStop}%
\bibitem [{\citenamefont {Heikkil{\"a}}\ \emph {et~al.}(2011)\citenamefont
  {Heikkil{\"a}}, \citenamefont {Kopnin},\ and\ \citenamefont
  {Volovik}}]{Heikkila:2011}%
  \BibitemOpen
  \bibfield  {author} {\bibinfo {author} {\bibfnamefont {T.~T.}\ \bibnamefont
  {Heikkil{\"a}}}, \bibinfo {author} {\bibfnamefont {N.~B.}\ \bibnamefont
  {Kopnin}},\ and\ \bibinfo {author} {\bibfnamefont {G.~E.}\ \bibnamefont
  {Volovik}},\ }\bibfield  {title} {\bibinfo {title} {Flat bands in topological
  media},\ }\href {https://doi.org/10.1134/S0021364011150045} {\bibfield
  {journal} {\bibinfo  {journal} {JETP Letters}\ }\textbf {\bibinfo {volume}
  {94}},\ \bibinfo {pages} {233} (\bibinfo {year} {2011})}\BibitemShut
  {NoStop}%
\bibitem [{\citenamefont {Derzhko}\ \emph {et~al.}(2015)\citenamefont
  {Derzhko}, \citenamefont {Richter},\ and\ \citenamefont
  {Maksymenko}}]{doi:10.1142/S0217979215300078}%
  \BibitemOpen
  \bibfield  {author} {\bibinfo {author} {\bibfnamefont {O.}~\bibnamefont
  {Derzhko}}, \bibinfo {author} {\bibfnamefont {J.}~\bibnamefont {Richter}},\
  and\ \bibinfo {author} {\bibfnamefont {M.}~\bibnamefont {Maksymenko}},\
  }\bibfield  {title} {\bibinfo {title} {Strongly correlated flat-band systems:
  The route from heisenberg spins to hubbard electrons},\ }\href
  {https://doi.org/10.1142/S0217979215300078} {\bibfield  {journal} {\bibinfo
  {journal} {International Journal of Modern Physics B}\ }\textbf {\bibinfo
  {volume} {29}},\ \bibinfo {pages} {1530007} (\bibinfo {year} {2015})},\
  \Eprint {https://arxiv.org/abs/https://doi.org/10.1142/S0217979215300078}
  {https://doi.org/10.1142/S0217979215300078} \BibitemShut {NoStop}%
\bibitem [{\citenamefont {Peotta}\ and\ \citenamefont
  {T{\"o}rm{\"a}}(2015)}]{Peotta2015}%
  \BibitemOpen
  \bibfield  {author} {\bibinfo {author} {\bibfnamefont {S.}~\bibnamefont
  {Peotta}}\ and\ \bibinfo {author} {\bibfnamefont {P.}~\bibnamefont
  {T{\"o}rm{\"a}}},\ }\bibfield  {title} {\bibinfo {title} {Superfluidity in
  topologically nontrivial flat bands},\ }\href
  {https://doi.org/10.1038/ncomms9944} {\bibfield  {journal} {\bibinfo
  {journal} {Nature Communications}\ }\textbf {\bibinfo {volume} {6}},\
  \bibinfo {pages} {8944} (\bibinfo {year} {2015})}\BibitemShut {NoStop}%
\bibitem [{\citenamefont {Noda}\ \emph {et~al.}(2014)\citenamefont {Noda},
  \citenamefont {Inaba},\ and\ \citenamefont {Yamashita}}]{PhysRevA.90.043624}%
  \BibitemOpen
  \bibfield  {author} {\bibinfo {author} {\bibfnamefont {K.}~\bibnamefont
  {Noda}}, \bibinfo {author} {\bibfnamefont {K.}~\bibnamefont {Inaba}},\ and\
  \bibinfo {author} {\bibfnamefont {M.}~\bibnamefont {Yamashita}},\ }\bibfield
  {title} {\bibinfo {title} {Flat-band ferromagnetism in the multilayer lieb
  optical lattice},\ }\href {https://doi.org/10.1103/PhysRevA.90.043624}
  {\bibfield  {journal} {\bibinfo  {journal} {Phys. Rev. A}\ }\textbf {\bibinfo
  {volume} {90}},\ \bibinfo {pages} {043624} (\bibinfo {year}
  {2014})}\BibitemShut {NoStop}%
\bibitem [{\citenamefont {Cao}\ \emph {et~al.}(2018{\natexlab{b}})\citenamefont
  {Cao}, \citenamefont {Fatemi}, \citenamefont {Demir}, \citenamefont {Fang},
  \citenamefont {Tomarken}, \citenamefont {Luo}, \citenamefont
  {Sanchez-Yamagishi}, \citenamefont {Watanabe}, \citenamefont {Taniguchi},
  \citenamefont {Kaxiras}, \citenamefont {Ashoori},\ and\ \citenamefont
  {Jarillo-Herrero}}]{Cao2018}%
  \BibitemOpen
  \bibfield  {author} {\bibinfo {author} {\bibfnamefont {Y.}~\bibnamefont
  {Cao}}, \bibinfo {author} {\bibfnamefont {V.}~\bibnamefont {Fatemi}},
  \bibinfo {author} {\bibfnamefont {A.}~\bibnamefont {Demir}}, \bibinfo
  {author} {\bibfnamefont {S.}~\bibnamefont {Fang}}, \bibinfo {author}
  {\bibfnamefont {S.~L.}\ \bibnamefont {Tomarken}}, \bibinfo {author}
  {\bibfnamefont {J.~Y.}\ \bibnamefont {Luo}}, \bibinfo {author} {\bibfnamefont
  {J.~D.}\ \bibnamefont {Sanchez-Yamagishi}}, \bibinfo {author} {\bibfnamefont
  {K.}~\bibnamefont {Watanabe}}, \bibinfo {author} {\bibfnamefont
  {T.}~\bibnamefont {Taniguchi}}, \bibinfo {author} {\bibfnamefont
  {E.}~\bibnamefont {Kaxiras}}, \bibinfo {author} {\bibfnamefont {R.~C.}\
  \bibnamefont {Ashoori}},\ and\ \bibinfo {author} {\bibfnamefont
  {P.}~\bibnamefont {Jarillo-Herrero}},\ }\bibfield  {title} {\bibinfo {title}
  {Correlated insulator behaviour at half-filling in magic-angle graphene
  superlattices},\ }\href {https://doi.org/10.1038/nature26154} {\bibfield
  {journal} {\bibinfo  {journal} {Nature}\ }\textbf {\bibinfo {volume} {556}},\
  \bibinfo {pages} {80 EP } (\bibinfo {year} {2018}{\natexlab{b}})}\BibitemShut
  {NoStop}%
\bibitem [{\citenamefont {Cao}\ \emph {et~al.}(2018{\natexlab{c}})\citenamefont
  {Cao}, \citenamefont {Fatemi}, \citenamefont {Fang}, \citenamefont
  {Watanabe}, \citenamefont {Taniguchi}, \citenamefont {Kaxiras},\ and\
  \citenamefont {Jarillo-Herrero}}]{Cao20181}%
  \BibitemOpen
  \bibfield  {author} {\bibinfo {author} {\bibfnamefont {Y.}~\bibnamefont
  {Cao}}, \bibinfo {author} {\bibfnamefont {V.}~\bibnamefont {Fatemi}},
  \bibinfo {author} {\bibfnamefont {S.}~\bibnamefont {Fang}}, \bibinfo {author}
  {\bibfnamefont {K.}~\bibnamefont {Watanabe}}, \bibinfo {author}
  {\bibfnamefont {T.}~\bibnamefont {Taniguchi}}, \bibinfo {author}
  {\bibfnamefont {E.}~\bibnamefont {Kaxiras}},\ and\ \bibinfo {author}
  {\bibfnamefont {P.}~\bibnamefont {Jarillo-Herrero}},\ }\bibfield  {title}
  {\bibinfo {title} {Unconventional superconductivity in magic-angle graphene
  superlattices},\ }\href {https://doi.org/10.1038/nature26160} {\bibfield
  {journal} {\bibinfo  {journal} {Nature}\ }\textbf {\bibinfo {volume} {556}},\
  \bibinfo {pages} {43 EP } (\bibinfo {year} {2018}{\natexlab{c}})}\BibitemShut
  {NoStop}%
\bibitem [{\citenamefont {Yankowitz}\ \emph {et~al.}(2019)\citenamefont
  {Yankowitz}, \citenamefont {Chen}, \citenamefont {Polshyn}, \citenamefont
  {Zhang}, \citenamefont {Watanabe}, \citenamefont {Taniguchi}, \citenamefont
  {Graf}, \citenamefont {Young},\ and\ \citenamefont {Dean}}]{Yankowitz1059}%
  \BibitemOpen
  \bibfield  {author} {\bibinfo {author} {\bibfnamefont {M.}~\bibnamefont
  {Yankowitz}}, \bibinfo {author} {\bibfnamefont {S.}~\bibnamefont {Chen}},
  \bibinfo {author} {\bibfnamefont {H.}~\bibnamefont {Polshyn}}, \bibinfo
  {author} {\bibfnamefont {Y.}~\bibnamefont {Zhang}}, \bibinfo {author}
  {\bibfnamefont {K.}~\bibnamefont {Watanabe}}, \bibinfo {author}
  {\bibfnamefont {T.}~\bibnamefont {Taniguchi}}, \bibinfo {author}
  {\bibfnamefont {D.}~\bibnamefont {Graf}}, \bibinfo {author} {\bibfnamefont
  {A.~F.}\ \bibnamefont {Young}},\ and\ \bibinfo {author} {\bibfnamefont
  {C.~R.}\ \bibnamefont {Dean}},\ }\bibfield  {title} {\bibinfo {title} {Tuning
  superconductivity in twisted bilayer graphene},\ }\href
  {https://doi.org/10.1126/science.aav1910} {\bibfield  {journal} {\bibinfo
  {journal} {Science}\ }\textbf {\bibinfo {volume} {363}},\ \bibinfo {pages}
  {1059} (\bibinfo {year} {2019})}\BibitemShut {NoStop}%
\bibitem [{\citenamefont {Tran}\ and\ \citenamefont
  {Nguyen}(2018)}]{PhysRevB.97.155125}%
  \BibitemOpen
  \bibfield  {author} {\bibinfo {author} {\bibfnamefont {M.-T.}\ \bibnamefont
  {Tran}}\ and\ \bibinfo {author} {\bibfnamefont {T.~T.}\ \bibnamefont
  {Nguyen}},\ }\bibfield  {title} {\bibinfo {title} {Molecular kondo effect in
  flat-band lattices},\ }\href {https://doi.org/10.1103/PhysRevB.97.155125}
  {\bibfield  {journal} {\bibinfo  {journal} {Phys. Rev. B}\ }\textbf {\bibinfo
  {volume} {97}},\ \bibinfo {pages} {155125} (\bibinfo {year}
  {2018})}\BibitemShut {NoStop}%
\bibitem [{\citenamefont {Ramires}\ and\ \citenamefont
  {Lado}(2021)}]{PhysRevLett.127.026401}%
  \BibitemOpen
  \bibfield  {author} {\bibinfo {author} {\bibfnamefont {A.}~\bibnamefont
  {Ramires}}\ and\ \bibinfo {author} {\bibfnamefont {J.~L.}\ \bibnamefont
  {Lado}},\ }\bibfield  {title} {\bibinfo {title} {Emulating heavy fermions in
  twisted trilayer graphene},\ }\href
  {https://doi.org/10.1103/PhysRevLett.127.026401} {\bibfield  {journal}
  {\bibinfo  {journal} {Phys. Rev. Lett.}\ }\textbf {\bibinfo {volume} {127}},\
  \bibinfo {pages} {026401} (\bibinfo {year} {2021})}\BibitemShut {NoStop}%
\bibitem [{\citenamefont {Jiang}\ \emph {et~al.}(2019)\citenamefont {Jiang},
  \citenamefont {Lai}, \citenamefont {Watanabe}, \citenamefont {Taniguchi},
  \citenamefont {Haule}, \citenamefont {Mao},\ and\ \citenamefont
  {Andrei}}]{Jiang2019}%
  \BibitemOpen
  \bibfield  {author} {\bibinfo {author} {\bibfnamefont {Y.}~\bibnamefont
  {Jiang}}, \bibinfo {author} {\bibfnamefont {X.}~\bibnamefont {Lai}}, \bibinfo
  {author} {\bibfnamefont {K.}~\bibnamefont {Watanabe}}, \bibinfo {author}
  {\bibfnamefont {T.}~\bibnamefont {Taniguchi}}, \bibinfo {author}
  {\bibfnamefont {K.}~\bibnamefont {Haule}}, \bibinfo {author} {\bibfnamefont
  {J.}~\bibnamefont {Mao}},\ and\ \bibinfo {author} {\bibfnamefont {E.~Y.}\
  \bibnamefont {Andrei}},\ }\bibfield  {title} {\bibinfo {title} {Charge order
  and broken rotational symmetry in magic-angle twisted bilayer graphene},\
  }\href {https://doi.org/10.1038/s41586-019-1460-4} {\bibfield  {journal}
  {\bibinfo  {journal} {Nature}\ }\textbf {\bibinfo {volume} {573}},\ \bibinfo
  {pages} {91} (\bibinfo {year} {2019})}\BibitemShut {NoStop}%
\bibitem [{\citenamefont {Xu}\ \emph {et~al.}(2018)\citenamefont {Xu},
  \citenamefont {Law},\ and\ \citenamefont {Lee}}]{PhysRevB.98.121406}%
  \BibitemOpen
  \bibfield  {author} {\bibinfo {author} {\bibfnamefont {X.~Y.}\ \bibnamefont
  {Xu}}, \bibinfo {author} {\bibfnamefont {K.~T.}\ \bibnamefont {Law}},\ and\
  \bibinfo {author} {\bibfnamefont {P.~A.}\ \bibnamefont {Lee}},\ }\bibfield
  {title} {\bibinfo {title} {Kekul\'e valence bond order in an extended hubbard
  model on the honeycomb lattice with possible applications to twisted bilayer
  graphene},\ }\href {https://doi.org/10.1103/PhysRevB.98.121406} {\bibfield
  {journal} {\bibinfo  {journal} {Phys. Rev. B}\ }\textbf {\bibinfo {volume}
  {98}},\ \bibinfo {pages} {121406} (\bibinfo {year} {2018})}\BibitemShut
  {NoStop}%
\bibitem [{\citenamefont {Kumar}\ \emph {et~al.}(2019)\citenamefont {Kumar},
  \citenamefont {Vanhala},\ and\ \citenamefont {T\"orm\"a}}]{Kumar2019}%
  \BibitemOpen
  \bibfield  {author} {\bibinfo {author} {\bibfnamefont {P.}~\bibnamefont
  {Kumar}}, \bibinfo {author} {\bibfnamefont {T.~I.}\ \bibnamefont {Vanhala}},\
  and\ \bibinfo {author} {\bibfnamefont {P.}~\bibnamefont {T\"orm\"a}},\
  }\bibfield  {title} {\bibinfo {title} {Magnetization, $d$-wave
  superconductivity, and non-{F}ermi-liquid behavior in a crossover from
  dispersive to flat bands},\ }\href
  {https://doi.org/10.1103/PhysRevB.100.125141} {\bibfield  {journal} {\bibinfo
   {journal} {Phys. Rev. B}\ }\textbf {\bibinfo {volume} {100}},\ \bibinfo
  {pages} {125141} (\bibinfo {year} {2019})}\BibitemShut {NoStop}%
\bibitem [{\citenamefont {Kumar}\ \emph {et~al.}(2021)\citenamefont {Kumar},
  \citenamefont {Peotta}, \citenamefont {Takasu}, \citenamefont {Takahashi},\
  and\ \citenamefont {T\"orm\"a}}]{PhysRevA.103.L031301}%
  \BibitemOpen
  \bibfield  {author} {\bibinfo {author} {\bibfnamefont {P.}~\bibnamefont
  {Kumar}}, \bibinfo {author} {\bibfnamefont {S.}~\bibnamefont {Peotta}},
  \bibinfo {author} {\bibfnamefont {Y.}~\bibnamefont {Takasu}}, \bibinfo
  {author} {\bibfnamefont {Y.}~\bibnamefont {Takahashi}},\ and\ \bibinfo
  {author} {\bibfnamefont {P.}~\bibnamefont {T\"orm\"a}},\ }\bibfield  {title}
  {\bibinfo {title} {Flat-band-induced non-fermi-liquid behavior of
  multicomponent fermions},\ }\href
  {https://doi.org/10.1103/PhysRevA.103.L031301} {\bibfield  {journal}
  {\bibinfo  {journal} {Phys. Rev. A}\ }\textbf {\bibinfo {volume} {103}},\
  \bibinfo {pages} {L031301} (\bibinfo {year} {2021})}\BibitemShut {NoStop}%
\bibitem [{\citenamefont {Coleman}(1984)}]{PhysRevB.29.3035}%
  \BibitemOpen
  \bibfield  {author} {\bibinfo {author} {\bibfnamefont {P.}~\bibnamefont
  {Coleman}},\ }\bibfield  {title} {\bibinfo {title} {New approach to the
  mixed-valence problem},\ }\href {https://doi.org/10.1103/PhysRevB.29.3035}
  {\bibfield  {journal} {\bibinfo  {journal} {Phys. Rev. B}\ }\textbf {\bibinfo
  {volume} {29}},\ \bibinfo {pages} {3035} (\bibinfo {year}
  {1984})}\BibitemShut {NoStop}%
\bibitem [{\citenamefont {Stewart}(2001)}]{RevModPhys.73.797}%
  \BibitemOpen
  \bibfield  {author} {\bibinfo {author} {\bibfnamefont {G.~R.}\ \bibnamefont
  {Stewart}},\ }\bibfield  {title} {\bibinfo {title} {Non-fermi-liquid behavior
  in $d$- and $f$-electron metals},\ }\href
  {https://doi.org/10.1103/RevModPhys.73.797} {\bibfield  {journal} {\bibinfo
  {journal} {Rev. Mod. Phys.}\ }\textbf {\bibinfo {volume} {73}},\ \bibinfo
  {pages} {797} (\bibinfo {year} {2001})}\BibitemShut {NoStop}%
\bibitem [{\citenamefont {Wirth}\ and\ \citenamefont
  {Steglich}(2016)}]{Wirth2016}%
  \BibitemOpen
  \bibfield  {author} {\bibinfo {author} {\bibfnamefont {S.}~\bibnamefont
  {Wirth}}\ and\ \bibinfo {author} {\bibfnamefont {F.}~\bibnamefont
  {Steglich}},\ }\bibfield  {title} {\bibinfo {title} {Exploring heavy fermions
  from macroscopic to microscopic length scales},\ }\bibfield  {journal}
  {\bibinfo  {journal} {Nature Reviews Materials}\ }\textbf {\bibinfo {volume}
  {1}},\ \href {https://doi.org/10.1038/natrevmats.2016.51}
  {10.1038/natrevmats.2016.51} (\bibinfo {year} {2016})\BibitemShut {NoStop}%
\bibitem [{\citenamefont {Tsunetsugu}\ \emph {et~al.}(1997)\citenamefont
  {Tsunetsugu}, \citenamefont {Sigrist},\ and\ \citenamefont
  {Ueda}}]{RevModPhys.69.809}%
  \BibitemOpen
  \bibfield  {author} {\bibinfo {author} {\bibfnamefont {H.}~\bibnamefont
  {Tsunetsugu}}, \bibinfo {author} {\bibfnamefont {M.}~\bibnamefont
  {Sigrist}},\ and\ \bibinfo {author} {\bibfnamefont {K.}~\bibnamefont
  {Ueda}},\ }\bibfield  {title} {\bibinfo {title} {The ground-state phase
  diagram of the one-dimensional kondo lattice model},\ }\href
  {https://doi.org/10.1103/RevModPhys.69.809} {\bibfield  {journal} {\bibinfo
  {journal} {Rev. Mod. Phys.}\ }\textbf {\bibinfo {volume} {69}},\ \bibinfo
  {pages} {809} (\bibinfo {year} {1997})}\BibitemShut {NoStop}%
\bibitem [{\citenamefont {Si}\ and\ \citenamefont {Steglich}(2010)}]{Si1161}%
  \BibitemOpen
  \bibfield  {author} {\bibinfo {author} {\bibfnamefont {Q.}~\bibnamefont
  {Si}}\ and\ \bibinfo {author} {\bibfnamefont {F.}~\bibnamefont {Steglich}},\
  }\bibfield  {title} {\bibinfo {title} {Heavy fermions and quantum phase
  transitions},\ }\href {https://doi.org/10.1126/science.1191195} {\bibfield
  {journal} {\bibinfo  {journal} {Science}\ }\textbf {\bibinfo {volume}
  {329}},\ \bibinfo {pages} {1161} (\bibinfo {year} {2010})}\BibitemShut
  {NoStop}%
\bibitem [{\citenamefont {Coleman}(2015)}]{coleman2015heavy}%
  \BibitemOpen
  \bibfield  {author} {\bibinfo {author} {\bibfnamefont {P.}~\bibnamefont
  {Coleman}},\ }\bibfield  {title} {\bibinfo {title} {Heavy fermions and the
  kondo lattice: a 21st century perspective},\ }\href@noop {} {\bibfield
  {journal} {\bibinfo  {journal} {arXiv preprint arXiv:1509.05769}\ } (\bibinfo
  {year} {2015})}\BibitemShut {NoStop}%
\bibitem [{\citenamefont {Ramires}\ and\ \citenamefont
  {Coleman}(2014)}]{PhysRevLett.112.116405}%
  \BibitemOpen
  \bibfield  {author} {\bibinfo {author} {\bibfnamefont {A.}~\bibnamefont
  {Ramires}}\ and\ \bibinfo {author} {\bibfnamefont {P.}~\bibnamefont
  {Coleman}},\ }\bibfield  {title} {\bibinfo {title} {Theory of the electron
  spin resonance in the heavy fermion metal
  $\ensuremath{\beta}\text{\ensuremath{-}}{\mathrm{ybalb}}_{4}$},\ }\href
  {https://doi.org/10.1103/PhysRevLett.112.116405} {\bibfield  {journal}
  {\bibinfo  {journal} {Phys. Rev. Lett.}\ }\textbf {\bibinfo {volume} {112}},\
  \bibinfo {pages} {116405} (\bibinfo {year} {2014})}\BibitemShut {NoStop}%
\bibitem [{\citenamefont {Ramires}\ and\ \citenamefont
  {Coleman}(2016)}]{PhysRevB.93.035120}%
  \BibitemOpen
  \bibfield  {author} {\bibinfo {author} {\bibfnamefont {A.}~\bibnamefont
  {Ramires}}\ and\ \bibinfo {author} {\bibfnamefont {P.}~\bibnamefont
  {Coleman}},\ }\bibfield  {title} {\bibinfo {title} {Supersymmetric approach
  to heavy fermion systems},\ }\href
  {https://doi.org/10.1103/PhysRevB.93.035120} {\bibfield  {journal} {\bibinfo
  {journal} {Phys. Rev. B}\ }\textbf {\bibinfo {volume} {93}},\ \bibinfo
  {pages} {035120} (\bibinfo {year} {2016})}\BibitemShut {NoStop}%
\bibitem [{\citenamefont {{Va{\v{n}}o}}\ \emph {et~al.}(2021)\citenamefont
  {{Va{\v{n}}o}}, \citenamefont {{Amini}}, \citenamefont {{Ganguli}},
  \citenamefont {{Chen}}, \citenamefont {{Lado}}, \citenamefont
  {{Kezilebieke}},\ and\ \citenamefont {{Liljeroth}}}]{2021arXiv210311989V}%
  \BibitemOpen
  \bibfield  {author} {\bibinfo {author} {\bibfnamefont {V.}~\bibnamefont
  {{Va{\v{n}}o}}}, \bibinfo {author} {\bibfnamefont {M.}~\bibnamefont
  {{Amini}}}, \bibinfo {author} {\bibfnamefont {S.~C.}\ \bibnamefont
  {{Ganguli}}}, \bibinfo {author} {\bibfnamefont {G.}~\bibnamefont {{Chen}}},
  \bibinfo {author} {\bibfnamefont {J.~L.}\ \bibnamefont {{Lado}}}, \bibinfo
  {author} {\bibfnamefont {S.}~\bibnamefont {{Kezilebieke}}},\ and\ \bibinfo
  {author} {\bibfnamefont {P.}~\bibnamefont {{Liljeroth}}},\ }\bibfield
  {title} {\bibinfo {title} {{Artificial heavy fermions in a van der Waals
  heterostructure}},\ }\href@noop {} {\bibfield  {journal} {\bibinfo  {journal}
  {arXiv e-prints}\ ,\ \bibinfo {eid} {arXiv:2103.11989}} (\bibinfo {year}
  {2021})},\ \Eprint {https://arxiv.org/abs/2103.11989} {arXiv:2103.11989
  [cond-mat.mes-hall]} \BibitemShut {NoStop}%
\bibitem [{\citenamefont {Dzero}\ \emph {et~al.}(2010)\citenamefont {Dzero},
  \citenamefont {Sun}, \citenamefont {Galitski},\ and\ \citenamefont
  {Coleman}}]{PhysRevLett.104.106408}%
  \BibitemOpen
  \bibfield  {author} {\bibinfo {author} {\bibfnamefont {M.}~\bibnamefont
  {Dzero}}, \bibinfo {author} {\bibfnamefont {K.}~\bibnamefont {Sun}}, \bibinfo
  {author} {\bibfnamefont {V.}~\bibnamefont {Galitski}},\ and\ \bibinfo
  {author} {\bibfnamefont {P.}~\bibnamefont {Coleman}},\ }\bibfield  {title}
  {\bibinfo {title} {Topological kondo insulators},\ }\href
  {https://doi.org/10.1103/PhysRevLett.104.106408} {\bibfield  {journal}
  {\bibinfo  {journal} {Phys. Rev. Lett.}\ }\textbf {\bibinfo {volume} {104}},\
  \bibinfo {pages} {106408} (\bibinfo {year} {2010})}\BibitemShut {NoStop}%
\bibitem [{\citenamefont {Dzero}\ \emph {et~al.}(2016)\citenamefont {Dzero},
  \citenamefont {Xia}, \citenamefont {Galitski},\ and\ \citenamefont
  {Coleman}}]{Dzero2016}%
  \BibitemOpen
  \bibfield  {author} {\bibinfo {author} {\bibfnamefont {M.}~\bibnamefont
  {Dzero}}, \bibinfo {author} {\bibfnamefont {J.}~\bibnamefont {Xia}}, \bibinfo
  {author} {\bibfnamefont {V.}~\bibnamefont {Galitski}},\ and\ \bibinfo
  {author} {\bibfnamefont {P.}~\bibnamefont {Coleman}},\ }\bibfield  {title}
  {\bibinfo {title} {Topological kondo insulators},\ }\href
  {https://doi.org/10.1146/annurev-conmatphys-031214-014749} {\bibfield
  {journal} {\bibinfo  {journal} {Annual Review of Condensed Matter Physics}\
  }\textbf {\bibinfo {volume} {7}},\ \bibinfo {pages} {249} (\bibinfo {year}
  {2016})}\BibitemShut {NoStop}%
\bibitem [{\citenamefont {Legner}\ \emph {et~al.}(2015)\citenamefont {Legner},
  \citenamefont {R\"uegg},\ and\ \citenamefont
  {Sigrist}}]{PhysRevLett.115.156405}%
  \BibitemOpen
  \bibfield  {author} {\bibinfo {author} {\bibfnamefont {M.}~\bibnamefont
  {Legner}}, \bibinfo {author} {\bibfnamefont {A.}~\bibnamefont {R\"uegg}},\
  and\ \bibinfo {author} {\bibfnamefont {M.}~\bibnamefont {Sigrist}},\
  }\bibfield  {title} {\bibinfo {title} {Surface-state spin textures and mirror
  chern numbers in topological kondo insulators},\ }\href
  {https://doi.org/10.1103/PhysRevLett.115.156405} {\bibfield  {journal}
  {\bibinfo  {journal} {Phys. Rev. Lett.}\ }\textbf {\bibinfo {volume} {115}},\
  \bibinfo {pages} {156405} (\bibinfo {year} {2015})}\BibitemShut {NoStop}%
\bibitem [{\citenamefont {Jiao}\ \emph {et~al.}(2020)\citenamefont {Jiao},
  \citenamefont {Howard}, \citenamefont {Ran}, \citenamefont {Wang},
  \citenamefont {Rodriguez}, \citenamefont {Sigrist}, \citenamefont {Wang},
  \citenamefont {Butch},\ and\ \citenamefont {Madhavan}}]{Jiao2020}%
  \BibitemOpen
  \bibfield  {author} {\bibinfo {author} {\bibfnamefont {L.}~\bibnamefont
  {Jiao}}, \bibinfo {author} {\bibfnamefont {S.}~\bibnamefont {Howard}},
  \bibinfo {author} {\bibfnamefont {S.}~\bibnamefont {Ran}}, \bibinfo {author}
  {\bibfnamefont {Z.}~\bibnamefont {Wang}}, \bibinfo {author} {\bibfnamefont
  {J.~O.}\ \bibnamefont {Rodriguez}}, \bibinfo {author} {\bibfnamefont
  {M.}~\bibnamefont {Sigrist}}, \bibinfo {author} {\bibfnamefont
  {Z.}~\bibnamefont {Wang}}, \bibinfo {author} {\bibfnamefont {N.~P.}\
  \bibnamefont {Butch}},\ and\ \bibinfo {author} {\bibfnamefont
  {V.}~\bibnamefont {Madhavan}},\ }\bibfield  {title} {\bibinfo {title} {Chiral
  superconductivity in heavy-fermion metal {UTe}2},\ }\href
  {https://doi.org/10.1038/s41586-020-2122-2} {\bibfield  {journal} {\bibinfo
  {journal} {Nature}\ }\textbf {\bibinfo {volume} {579}},\ \bibinfo {pages}
  {523} (\bibinfo {year} {2020})}\BibitemShut {NoStop}%
\bibitem [{\citenamefont {Titvinidze}\ \emph {et~al.}(2014)\citenamefont
  {Titvinidze}, \citenamefont {Schwabe},\ and\ \citenamefont
  {Potthoff}}]{PhysRevB.90.045112}%
  \BibitemOpen
  \bibfield  {author} {\bibinfo {author} {\bibfnamefont {I.}~\bibnamefont
  {Titvinidze}}, \bibinfo {author} {\bibfnamefont {A.}~\bibnamefont
  {Schwabe}},\ and\ \bibinfo {author} {\bibfnamefont {M.}~\bibnamefont
  {Potthoff}},\ }\bibfield  {title} {\bibinfo {title} {Ferromagnetism of
  magnetic impurities coupled indirectly via conduction electrons: Insights
  from various theoretical approaches},\ }\href
  {https://doi.org/10.1103/PhysRevB.90.045112} {\bibfield  {journal} {\bibinfo
  {journal} {Phys. Rev. B}\ }\textbf {\bibinfo {volume} {90}},\ \bibinfo
  {pages} {045112} (\bibinfo {year} {2014})}\BibitemShut {NoStop}%
\bibitem [{\citenamefont {Schwabe}\ \emph {et~al.}(2013)\citenamefont
  {Schwabe}, \citenamefont {Titvinidze},\ and\ \citenamefont
  {Potthoff}}]{PhysRevB.88.121107}%
  \BibitemOpen
  \bibfield  {author} {\bibinfo {author} {\bibfnamefont {A.}~\bibnamefont
  {Schwabe}}, \bibinfo {author} {\bibfnamefont {I.}~\bibnamefont
  {Titvinidze}},\ and\ \bibinfo {author} {\bibfnamefont {M.}~\bibnamefont
  {Potthoff}},\ }\bibfield  {title} {\bibinfo {title} {Inverse indirect
  magnetic exchange},\ }\href {https://doi.org/10.1103/PhysRevB.88.121107}
  {\bibfield  {journal} {\bibinfo  {journal} {Phys. Rev. B}\ }\textbf {\bibinfo
  {volume} {88}},\ \bibinfo {pages} {121107} (\bibinfo {year}
  {2013})}\BibitemShut {NoStop}%
\bibitem [{\citenamefont {Costa}\ \emph {et~al.}(2018)\citenamefont {Costa},
  \citenamefont {Ara\'ujo}, \citenamefont {Lima}, \citenamefont {Paiva},
  \citenamefont {dos Santos},\ and\ \citenamefont
  {Scalettar}}]{PhysRevB.97.085123}%
  \BibitemOpen
  \bibfield  {author} {\bibinfo {author} {\bibfnamefont {N.~C.}\ \bibnamefont
  {Costa}}, \bibinfo {author} {\bibfnamefont {M.~V.}\ \bibnamefont {Ara\'ujo}},
  \bibinfo {author} {\bibfnamefont {J.~P.}\ \bibnamefont {Lima}}, \bibinfo
  {author} {\bibfnamefont {T.}~\bibnamefont {Paiva}}, \bibinfo {author}
  {\bibfnamefont {R.~R.}\ \bibnamefont {dos Santos}},\ and\ \bibinfo {author}
  {\bibfnamefont {R.~T.}\ \bibnamefont {Scalettar}},\ }\bibfield  {title}
  {\bibinfo {title} {Compressible ferrimagnetism in the depleted periodic
  anderson model},\ }\href {https://doi.org/10.1103/PhysRevB.97.085123}
  {\bibfield  {journal} {\bibinfo  {journal} {Phys. Rev. B}\ }\textbf {\bibinfo
  {volume} {97}},\ \bibinfo {pages} {085123} (\bibinfo {year}
  {2018})}\BibitemShut {NoStop}%
\bibitem [{\citenamefont {Ramires}\ and\ \citenamefont
  {Lado}(2019)}]{PhysRevB.99.245118}%
  \BibitemOpen
  \bibfield  {author} {\bibinfo {author} {\bibfnamefont {A.}~\bibnamefont
  {Ramires}}\ and\ \bibinfo {author} {\bibfnamefont {J.~L.}\ \bibnamefont
  {Lado}},\ }\bibfield  {title} {\bibinfo {title} {Impurity-induced triple
  point fermions in twisted bilayer graphene},\ }\href
  {https://doi.org/10.1103/PhysRevB.99.245118} {\bibfield  {journal} {\bibinfo
  {journal} {Phys. Rev. B}\ }\textbf {\bibinfo {volume} {99}},\ \bibinfo
  {pages} {245118} (\bibinfo {year} {2019})}\BibitemShut {NoStop}%
\bibitem [{\citenamefont {Lopez-Bezanilla}\ and\ \citenamefont
  {Lado}(2019)}]{PhysRevMaterials.3.084003}%
  \BibitemOpen
  \bibfield  {author} {\bibinfo {author} {\bibfnamefont {A.}~\bibnamefont
  {Lopez-Bezanilla}}\ and\ \bibinfo {author} {\bibfnamefont {J.~L.}\
  \bibnamefont {Lado}},\ }\bibfield  {title} {\bibinfo {title} {Defect-induced
  magnetism and yu-shiba-rusinov states in twisted bilayer graphene},\ }\href
  {https://doi.org/10.1103/PhysRevMaterials.3.084003} {\bibfield  {journal}
  {\bibinfo  {journal} {Phys. Rev. Materials}\ }\textbf {\bibinfo {volume}
  {3}},\ \bibinfo {pages} {084003} (\bibinfo {year} {2019})}\BibitemShut
  {NoStop}%
\bibitem [{\citenamefont {Ruderman}\ and\ \citenamefont
  {Kittel}(1954)}]{PhysRev.96.99}%
  \BibitemOpen
  \bibfield  {author} {\bibinfo {author} {\bibfnamefont {M.~A.}\ \bibnamefont
  {Ruderman}}\ and\ \bibinfo {author} {\bibfnamefont {C.}~\bibnamefont
  {Kittel}},\ }\bibfield  {title} {\bibinfo {title} {Indirect exchange coupling
  of nuclear magnetic moments by conduction electrons},\ }\href
  {https://doi.org/10.1103/PhysRev.96.99} {\bibfield  {journal} {\bibinfo
  {journal} {Phys. Rev.}\ }\textbf {\bibinfo {volume} {96}},\ \bibinfo {pages}
  {99} (\bibinfo {year} {1954})}\BibitemShut {NoStop}%
\bibitem [{\citenamefont {Yosida}(1957)}]{PhysRev.106.893}%
  \BibitemOpen
  \bibfield  {author} {\bibinfo {author} {\bibfnamefont {K.}~\bibnamefont
  {Yosida}},\ }\bibfield  {title} {\bibinfo {title} {Magnetic properties of
  cu-mn alloys},\ }\href {https://doi.org/10.1103/PhysRev.106.893} {\bibfield
  {journal} {\bibinfo  {journal} {Phys. Rev.}\ }\textbf {\bibinfo {volume}
  {106}},\ \bibinfo {pages} {893} (\bibinfo {year} {1957})}\BibitemShut
  {NoStop}%
\bibitem [{\citenamefont {Kasuya}(1956)}]{Kasuya1956}%
  \BibitemOpen
  \bibfield  {author} {\bibinfo {author} {\bibfnamefont {T.}~\bibnamefont
  {Kasuya}},\ }\bibfield  {title} {\bibinfo {title} {A theory of metallic
  ferro- and antiferromagnetism on zener{\textquotesingle}s model},\ }\href
  {https://doi.org/10.1143/ptp.16.45} {\bibfield  {journal} {\bibinfo
  {journal} {Progress of Theoretical Physics}\ }\textbf {\bibinfo {volume}
  {16}},\ \bibinfo {pages} {45} (\bibinfo {year} {1956})}\BibitemShut {NoStop}%
\bibitem [{\citenamefont {Sutherland}(1986)}]{PhysRevB.34.5208}%
  \BibitemOpen
  \bibfield  {author} {\bibinfo {author} {\bibfnamefont {B.}~\bibnamefont
  {Sutherland}},\ }\bibfield  {title} {\bibinfo {title} {Localization of
  electronic wave functions due to local topology},\ }\href
  {https://doi.org/10.1103/PhysRevB.34.5208} {\bibfield  {journal} {\bibinfo
  {journal} {Phys. Rev. B}\ }\textbf {\bibinfo {volume} {34}},\ \bibinfo
  {pages} {5208} (\bibinfo {year} {1986})}\BibitemShut {NoStop}%
\bibitem [{\citenamefont {Brouwer}\ \emph {et~al.}(2002)\citenamefont
  {Brouwer}, \citenamefont {Racine}, \citenamefont {Furusaki}, \citenamefont
  {Hatsugai}, \citenamefont {Morita},\ and\ \citenamefont
  {Mudry}}]{PhysRevB.66.014204}%
  \BibitemOpen
  \bibfield  {author} {\bibinfo {author} {\bibfnamefont {P.~W.}\ \bibnamefont
  {Brouwer}}, \bibinfo {author} {\bibfnamefont {E.}~\bibnamefont {Racine}},
  \bibinfo {author} {\bibfnamefont {A.}~\bibnamefont {Furusaki}}, \bibinfo
  {author} {\bibfnamefont {Y.}~\bibnamefont {Hatsugai}}, \bibinfo {author}
  {\bibfnamefont {Y.}~\bibnamefont {Morita}},\ and\ \bibinfo {author}
  {\bibfnamefont {C.}~\bibnamefont {Mudry}},\ }\bibfield  {title} {\bibinfo
  {title} {Zero modes in the random hopping model},\ }\href
  {https://doi.org/10.1103/PhysRevB.66.014204} {\bibfield  {journal} {\bibinfo
  {journal} {Phys. Rev. B}\ }\textbf {\bibinfo {volume} {66}},\ \bibinfo
  {pages} {014204} (\bibinfo {year} {2002})}\BibitemShut {NoStop}%
\bibitem [{\citenamefont {Mielke}(1991)}]{Mielke1991}%
  \BibitemOpen
  \bibfield  {author} {\bibinfo {author} {\bibfnamefont {A.}~\bibnamefont
  {Mielke}},\ }\bibfield  {title} {\bibinfo {title} {Ferromagnetism in the
  hubbard model on line graphs and further considerations},\ }\href
  {https://doi.org/10.1088/0305-4470/24/14/018} {\bibfield  {journal} {\bibinfo
   {journal} {Journal of Physics A: Mathematical and General}\ }\textbf
  {\bibinfo {volume} {24}},\ \bibinfo {pages} {3311} (\bibinfo {year}
  {1991})}\BibitemShut {NoStop}%
\bibitem [{\citenamefont {Lieb}(1989)}]{PhysRevLett.62.1201}%
  \BibitemOpen
  \bibfield  {author} {\bibinfo {author} {\bibfnamefont {E.~H.}\ \bibnamefont
  {Lieb}},\ }\bibfield  {title} {\bibinfo {title} {Two theorems on the hubbard
  model},\ }\href {https://doi.org/10.1103/PhysRevLett.62.1201} {\bibfield
  {journal} {\bibinfo  {journal} {Phys. Rev. Lett.}\ }\textbf {\bibinfo
  {volume} {62}},\ \bibinfo {pages} {1201} (\bibinfo {year}
  {1989})}\BibitemShut {NoStop}%
\bibitem [{\citenamefont {Georges}\ \emph {et~al.}(1996)\citenamefont
  {Georges}, \citenamefont {Kotliar}, \citenamefont {Krauth},\ and\
  \citenamefont {Rozenberg}}]{RevModPhys.68.13}%
  \BibitemOpen
  \bibfield  {author} {\bibinfo {author} {\bibfnamefont {A.}~\bibnamefont
  {Georges}}, \bibinfo {author} {\bibfnamefont {G.}~\bibnamefont {Kotliar}},
  \bibinfo {author} {\bibfnamefont {W.}~\bibnamefont {Krauth}},\ and\ \bibinfo
  {author} {\bibfnamefont {M.~J.}\ \bibnamefont {Rozenberg}},\ }\bibfield
  {title} {\bibinfo {title} {Dynamical mean-field theory of strongly correlated
  fermion systems and the limit of infinite dimensions},\ }\href
  {https://doi.org/10.1103/RevModPhys.68.13} {\bibfield  {journal} {\bibinfo
  {journal} {Rev. Mod. Phys.}\ }\textbf {\bibinfo {volume} {68}},\ \bibinfo
  {pages} {13} (\bibinfo {year} {1996})}\BibitemShut {NoStop}%
\bibitem [{\citenamefont {Assaad}\ and\ \citenamefont
  {Lang}(2007)}]{PhysRevB.76.035116}%
  \BibitemOpen
  \bibfield  {author} {\bibinfo {author} {\bibfnamefont {F.~F.}\ \bibnamefont
  {Assaad}}\ and\ \bibinfo {author} {\bibfnamefont {T.~C.}\ \bibnamefont
  {Lang}},\ }\bibfield  {title} {\bibinfo {title} {Diagrammatic determinantal
  quantum monte carlo methods: Projective schemes and applications to the
  hubbard-holstein model},\ }\href {https://doi.org/10.1103/PhysRevB.76.035116}
  {\bibfield  {journal} {\bibinfo  {journal} {Phys. Rev. B}\ }\textbf {\bibinfo
  {volume} {76}},\ \bibinfo {pages} {035116} (\bibinfo {year}
  {2007})}\BibitemShut {NoStop}%
\bibitem [{\citenamefont {Gull}(2008)}]{Gull}%
  \BibitemOpen
  \bibfield  {author} {\bibinfo {author} {\bibfnamefont {E.}~\bibnamefont
  {Gull}},\ }\emph {\bibinfo {title} {Continuous-Time Quantum Montecarlo
  Algorithms for Fermions}},\ \href@noop {} {Ph.D. thesis},\ \bibinfo  {school}
  {ETH ZURICH} (\bibinfo {year} {2008})\BibitemShut {NoStop}%
\bibitem [{\citenamefont {Vanhala}\ and\ \citenamefont
  {T\"orm\"a}(2018)}]{vanhala2017dynamical}%
  \BibitemOpen
  \bibfield  {author} {\bibinfo {author} {\bibfnamefont {T.~I.}\ \bibnamefont
  {Vanhala}}\ and\ \bibinfo {author} {\bibfnamefont {P.}~\bibnamefont
  {T\"orm\"a}},\ }\bibfield  {title} {\bibinfo {title} {Dynamical mean-field
  theory study of stripe order and $d$-wave superconductivity in the
  two-dimensional hubbard model},\ }\href
  {https://doi.org/10.1103/PhysRevB.97.075112} {\bibfield  {journal} {\bibinfo
  {journal} {Phys. Rev. B}\ }\textbf {\bibinfo {volume} {97}},\ \bibinfo
  {pages} {075112} (\bibinfo {year} {2018})}\BibitemShut {NoStop}%
\bibitem [{\citenamefont {Lichtenstein}\ and\ \citenamefont
  {Katsnelson}(2000)}]{PhysRevB.62.R9283}%
  \BibitemOpen
  \bibfield  {author} {\bibinfo {author} {\bibfnamefont {A.~I.}\ \bibnamefont
  {Lichtenstein}}\ and\ \bibinfo {author} {\bibfnamefont {M.~I.}\ \bibnamefont
  {Katsnelson}},\ }\bibfield  {title} {\bibinfo {title} {Antiferromagnetism and
  d-wave superconductivity in cuprates: A cluster dynamical mean-field
  theory},\ }\href {https://doi.org/10.1103/PhysRevB.62.R9283} {\bibfield
  {journal} {\bibinfo  {journal} {Phys. Rev. B}\ }\textbf {\bibinfo {volume}
  {62}},\ \bibinfo {pages} {R9283} (\bibinfo {year} {2000})}\BibitemShut
  {NoStop}%
\bibitem [{\citenamefont {Capone}\ and\ \citenamefont
  {Kotliar}(2006)}]{PhysRevB.74.054513}%
  \BibitemOpen
  \bibfield  {author} {\bibinfo {author} {\bibfnamefont {M.}~\bibnamefont
  {Capone}}\ and\ \bibinfo {author} {\bibfnamefont {G.}~\bibnamefont
  {Kotliar}},\ }\bibfield  {title} {\bibinfo {title} {Competition between
  $d$-wave superconductivity and antiferromagnetism in the two-dimensional
  hubbard model},\ }\href {https://doi.org/10.1103/PhysRevB.74.054513}
  {\bibfield  {journal} {\bibinfo  {journal} {Phys. Rev. B}\ }\textbf {\bibinfo
  {volume} {74}},\ \bibinfo {pages} {054513} (\bibinfo {year}
  {2006})}\BibitemShut {NoStop}%
\bibitem [{\citenamefont {Burdin}\ and\ \citenamefont
  {Fulde}(2007)}]{PhysRevB.76.104425}%
  \BibitemOpen
  \bibfield  {author} {\bibinfo {author} {\bibfnamefont {S.}~\bibnamefont
  {Burdin}}\ and\ \bibinfo {author} {\bibfnamefont {P.}~\bibnamefont {Fulde}},\
  }\bibfield  {title} {\bibinfo {title} {Random kondo alloys investigated with
  the coherent potential approximation},\ }\href
  {https://doi.org/10.1103/PhysRevB.76.104425} {\bibfield  {journal} {\bibinfo
  {journal} {Phys. Rev. B}\ }\textbf {\bibinfo {volume} {76}},\ \bibinfo
  {pages} {104425} (\bibinfo {year} {2007})}\BibitemShut {NoStop}%
\bibitem [{\citenamefont {Hirsch}(1985)}]{PhysRevB.31.4403}%
  \BibitemOpen
  \bibfield  {author} {\bibinfo {author} {\bibfnamefont {J.~E.}\ \bibnamefont
  {Hirsch}},\ }\bibfield  {title} {\bibinfo {title} {Two-dimensional hubbard
  model: Numerical simulation study},\ }\href
  {https://doi.org/10.1103/PhysRevB.31.4403} {\bibfield  {journal} {\bibinfo
  {journal} {Phys. Rev. B}\ }\textbf {\bibinfo {volume} {31}},\ \bibinfo
  {pages} {4403} (\bibinfo {year} {1985})}\BibitemShut {NoStop}%
\bibitem [{\citenamefont {Mermin}\ and\ \citenamefont
  {Wagner}(1966)}]{PhysRevLett.17.1133}%
  \BibitemOpen
  \bibfield  {author} {\bibinfo {author} {\bibfnamefont {N.~D.}\ \bibnamefont
  {Mermin}}\ and\ \bibinfo {author} {\bibfnamefont {H.}~\bibnamefont
  {Wagner}},\ }\bibfield  {title} {\bibinfo {title} {Absence of ferromagnetism
  or antiferromagnetism in one- or two-dimensional isotropic heisenberg
  models},\ }\href {https://doi.org/10.1103/PhysRevLett.17.1133} {\bibfield
  {journal} {\bibinfo  {journal} {Phys. Rev. Lett.}\ }\textbf {\bibinfo
  {volume} {17}},\ \bibinfo {pages} {1133} (\bibinfo {year}
  {1966})}\BibitemShut {NoStop}%
\bibitem [{\citenamefont {Hohenberg}(1967)}]{PhysRev.158.383}%
  \BibitemOpen
  \bibfield  {author} {\bibinfo {author} {\bibfnamefont {P.~C.}\ \bibnamefont
  {Hohenberg}},\ }\bibfield  {title} {\bibinfo {title} {Existence of long-range
  order in one and two dimensions},\ }\href
  {https://doi.org/10.1103/PhysRev.158.383} {\bibfield  {journal} {\bibinfo
  {journal} {Phys. Rev.}\ }\textbf {\bibinfo {volume} {158}},\ \bibinfo {pages}
  {383} (\bibinfo {year} {1967})}\BibitemShut {NoStop}%
\bibitem [{\citenamefont {Fratino}\ \emph {et~al.}(2017)\citenamefont
  {Fratino}, \citenamefont {S\'emon}, \citenamefont {Charlebois}, \citenamefont
  {Sordi},\ and\ \citenamefont {Tremblay}}]{PhysRevB.95.235109}%
  \BibitemOpen
  \bibfield  {author} {\bibinfo {author} {\bibfnamefont {L.}~\bibnamefont
  {Fratino}}, \bibinfo {author} {\bibfnamefont {P.}~\bibnamefont {S\'emon}},
  \bibinfo {author} {\bibfnamefont {M.}~\bibnamefont {Charlebois}}, \bibinfo
  {author} {\bibfnamefont {G.}~\bibnamefont {Sordi}},\ and\ \bibinfo {author}
  {\bibfnamefont {A.-M.~S.}\ \bibnamefont {Tremblay}},\ }\bibfield  {title}
  {\bibinfo {title} {Signatures of the mott transition in the antiferromagnetic
  state of the two-dimensional hubbard model},\ }\href
  {https://doi.org/10.1103/PhysRevB.95.235109} {\bibfield  {journal} {\bibinfo
  {journal} {Phys. Rev. B}\ }\textbf {\bibinfo {volume} {95}},\ \bibinfo
  {pages} {235109} (\bibinfo {year} {2017})}\BibitemShut {NoStop}%
\bibitem [{\citenamefont {White}(1992)}]{PhysRevLett.69.2863}%
  \BibitemOpen
  \bibfield  {author} {\bibinfo {author} {\bibfnamefont {S.~R.}\ \bibnamefont
  {White}},\ }\bibfield  {title} {\bibinfo {title} {Density matrix formulation
  for quantum renormalization groups},\ }\href
  {https://doi.org/10.1103/PhysRevLett.69.2863} {\bibfield  {journal} {\bibinfo
   {journal} {Phys. Rev. Lett.}\ }\textbf {\bibinfo {volume} {69}},\ \bibinfo
  {pages} {2863} (\bibinfo {year} {1992})}\BibitemShut {NoStop}%
\bibitem [{\citenamefont {Fishman}\ \emph {et~al.}()\citenamefont {Fishman},
  \citenamefont {White},\ and\ \citenamefont
  {Stoudenmire}}]{2020arXiv200714822F}%
  \BibitemOpen
  \bibfield  {author} {\bibinfo {author} {\bibfnamefont {M.}~\bibnamefont
  {Fishman}}, \bibinfo {author} {\bibfnamefont {S.~R.}\ \bibnamefont {White}},\
  and\ \bibinfo {author} {\bibfnamefont {E.~M.}\ \bibnamefont {Stoudenmire}},\
  }\href@noop {} {\bibinfo {title} {{The ITensor Software Library for Tensor
  Network Calculations}}},\ \Eprint {https://arxiv.org/abs/2007.14822}
  {arXiv:2007.14822 [cs.MS]} \BibitemShut {NoStop}%
\bibitem [{ITe()}]{ITensor}%
  \BibitemOpen
  \href@noop {} {\bibinfo  {journal} {\mbox{ITensor Library}
  http://itensor.org}\ }\BibitemShut {NoStop}%
\bibitem [{dmr()}]{dmrgpy}%
  \BibitemOpen
\bibfield  {journal} {  }\href@noop {} {\bibinfo  {journal} {\mbox{DMRGpy
  Library} https://github.com/joselado/dmrgpy}\ }\BibitemShut {NoStop}%
\end{thebibliography}%
\end{document}